\documentclass[aps,pra,twocolumn,showpacs,superscriptaddress]{revtex4-1}
\usepackage{amsmath,bm}
\usepackage{graphicx,color,braket,amssymb,amsfonts,eufrak}
\usepackage{epstopdf}
\usepackage[normalem]{ulem}

\begin{document}

\title{Attraction-induced dynamical stability of a Bose-Einstein condensate in a nonlinear lattice}

\author{Raka Dasgupta}\email{dasguptaraka@gmail.com}
\affiliation{Department of Physics, University of Calcutta, 92 Acharya Prafulla Chandra Road, Kolkata-700009, India}
\affiliation{Asia Pacific Center for Theoretical Physics (APCTP), Pohang, Gyeongbuk 37673, Korea}

\author{B. Prasanna Venkatesh}
\affiliation{Institute for Quantum Optics and Quantum Information of the Austrian Academy of Sciences, Technikerstra\ss e 21a, Innsbruck 6020, Austria}
\affiliation{Institute for Theoretical Physics, University of Innsbruck, A-6020 Innsbruck, Austria}
\affiliation{Asia Pacific Center for Theoretical Physics (APCTP), Pohang, Gyeongbuk 37673, Korea}

\author{Gentaro Watanabe}\email{gentaro@zju.edu.cn}
\affiliation{Department of Physics and Zhejiang Institute of Modern Physics, Zhejiang University, 38 Zheda Road, Hangzhou, Zhejiang 310027, China}
\affiliation{Center for Theoretical Physics of Complex Systems, Institute for Basic Science (IBS), Daejeon 34051, Korea}
\affiliation{University of Science and Technology (UST), Daejeon 34113, Korea}
\affiliation{Asia Pacific Center for Theoretical Physics (APCTP), Pohang, Gyeongbuk 37673, Korea}
\affiliation{Department of Physics, POSTECH, Pohang, Gyeongbuk 37673, Korea}

\begin{abstract}
We study multiple-period Bloch states of a Bose-Einstein condensate with spatially periodic interactomic interaction. Solving the Gross-Pitaevskii equation for the continuum model, and also using a simplified discrete version of it, we investigate the energy-band structures and the corresponding stability properties. We observe a new ``attraction-induced dynamical stability'' mechanism caused by the localization of the density distribution in the attractive domains of the system and the isolation of these higher-density regions. This makes the superfluid stable near the zone boundary, and also enhances the stability of higher-periodic states if the nonlinear interaction strength is sufficiently high.
\end{abstract}


\maketitle

\section{Introduction}

The study of nonlinear phenomena in Bose-Einstein condensates (BECs) of cold atomic gases has become a subject of immense interest, both from theoretical and experimental perspectives \cite{pethickbook, kevrebook, morsch_rmp}. Confining magnetic traps and/or optical lattices provide controllable externally applied potentials for a dilute BEC and appear as a linear term in the Gross-Pitaevskii (GP) equation governing the statics and dynamics of the order parameter. Further, the interatomic interactions lead to an atomic density dependent nonlinear term within the GP framework. The strength of this nonlinear term can be controlled by varying the scattering length via  magnetic \cite{inouye, court,roberts, moerd, timmer} or optical \cite{fedichev, bohn, fatemi, theis} Feshbach resonances \cite{bloch_rmp, chin_rmp}.

One intriguing aspect of cold atomic gases confined in optical lattices is the competition between the linear terms coming from the optical lattice and the nonlinear terms \cite{trombe1, bronski, kevre2} that allows for solitonic solutions \cite{burger, denschlag, khaykovich}, loop structures in the energy bands \cite{wu02, diakonov02, mueller02, pethick2, carr1, watanabe, sarma}, period doubling \cite{carr1, pethick1, yoon}, etc., and also gives rise to dynamical instabilities \cite{wu2, pethick2, wu03, modugno04, desarlo05, pethickbook}. Along this research direction, recently another interesting possibility has opened up where one may imagine having no linear periodic component at all (apart from the kinetic energy) in the GP equation but instead introducing periodicity in the system via a spatially periodic nonlinearity.
Such a system is termed as a ``nonlinear lattice'' \cite{sakaguchi, malomed_rmp, wu1}. Here both the nonlinearity and the periodicity are generated by a single term. Experimentally it has been realized with optical Feshbach resonances, by means of pulsed optical standing waves \cite{takahashi}.

A BEC with a spatially modulated interaction within a mean-field approximation is well described by the GP equation in one dimension (1D):
\begin{equation}
\label{GP1}
i\hbar \dfrac{\partial \psi}{\partial t}=-\dfrac{\hbar^2}{2m}\dfrac{\partial^2}{\partial x^2}\psi+ (V_1+V_2 \mbox{cos}2k_0x)|\psi|^2\psi\, ,
\end{equation}
which is valid when the average number of particles per site is much larger than unity, and density and temperature are sufficiently low so that the normal component is negligible.
Here the nonlinear term comprises a constant and a periodically modulated component. It is assumed that both $V_1$ and $V_2$ are positive quantities that can be controlled experimentally.  $k_0$ is connected to the period $d$ of the modulation by $k_0=\pi/d$ and it, in fact, is the wave number of the laser beam for the optical Feshbach resonance. $m$ is the mass of bosons and $\psi$ is the condensate wave function. In a recent work, the band structure and stability of this system were studied \cite{wu1}, considering the Bloch wave solutions for the lowest-energy bands.

We study the same system but go beyond the usual Bloch states (we call them period-1 solutions) that have the same periodicity as that of the modulated interaction. It is known that for BECs in a periodic potential, in addition to the conventional Bloch states, stationary states with periods twice, or even higher multiples of the lattice period emerge as well \cite{pethick1}.
Furthermore, these higher period states are shown to be energetically and dynamically stable in other systems like BECs with dipole-dipole interactions  in optical lattices \cite{maluckov}.
In the present work, for the case of periodically modulated interaction, we explore the possibility of having period-doubled stationary states (termed as period-2 solutions). Moreover we make a comparison between the stability regions of period-1 \cite{wu1} and period-2 energy bands.
We find that while the stability of the period-1 solutions can be qualitatively explained in terms of the overall averaged interaction as described in earlier studies \cite{wu1}, the stability of period-2 solutions demands for a more careful study of the dynamics of the system. We show that, in the period-2 case, BECs localized at each cell are more isolated and such isolation can stabilize the dynamics of the system, giving the central result of this paper: attraction-induced dynamical stability.

The paper is organized as follows. In Sec.~\ref{sec:discrete}, the system is described using a discrete model to obtain a basic sketch of the energy bands and the overall stability trends. In Sec.~\ref{sec:continuum} we deal with the full continuum model for the system, and solve the GP equation to study the band structures and the stability conditions. In Sec.~\ref{sec:mech} the stability mechanism is explained from a physical standpoint. We summarize the results in Sec.~\ref{sec:summary}.

\section{The discrete model \label{sec:discrete}}

\subsection{Formalism}

We first consider a simplified version of the system, where the uniform component of interaction is set to zero ($V_1=0$ and $V_2 \ne 0$), and map it in a discrete model \cite{pethick2, trombe2}. This is analogous to  an optical lattice in 1D. We reduce the system with a spatially periodic interaction in the continuum representation to a discrete representation by sampling just two points per period of the interaction (the maxima and minima of the interaction). Thus in this discrete model, the spacing between two sites is given by $\tilde{d}$ with the period of the interaction (i.e., the period of the original nonlinear lattice) $d = 2 \tilde{d}$. In this representation, the on-site interaction parameter alternates between $U$ and $-U$ at the adjacent sites.
To obtain periodic solutions, we can define a ``supercell'' that consists of two sites, with the lattice constant $d$. If instead of regular Bloch solutions, we consider a $p$-periodic solution, the length of the supercell will be $pd$, containing $2p$ \emph{discrete} lattice sites. 

A simple Hamiltonian for such a discrete model describing tunneling and interaction in this situation can be written as \cite{pethick2,trombe2}
\begin{equation}
\begin{split}
  H =& -K \sum_j (\psi_j^* \psi_{j+1} + \psi_{j+1}^* \psi_j) \\
& + \frac{U}{2} \left[\, \sum_{j={\rm even}}|\psi_j|^4 - \sum_{j={\rm odd}} |\psi_j|^4\, \right]\, ,
\label{hamilt}
\end{split}
\end{equation}
where $\psi_j$ is the amplitude at site $j$.
Here the first term in the equation signifies hopping between the nearest-neighbour sites characterized by the hopping parameter $K$, and the next term denotes the on-site inter-particle interaction. It is assumed that the odd-numbered sites are attractive, while the even-numbered sites are repulsive.

We aim to find stationary states with a fixed total number of particles. These are obtained by demanding that the variation of $H-\mu N$ ($\mu$ being the chemical potential) with respect to $\psi_j^*$ be zero. That is, 
\begin{equation}
\label{var1}
\begin{split}
U|\psi_j|^2\psi_j- K(\psi_{j+1}+\psi_{j-1})-\mu \psi_j=0\quad  (\mbox{for even \textit{j}}),\\
-U|\psi_j|^2\psi_j- K(\psi_{j+1}+\psi_{j-1})-\mu \psi_j=0\quad  (\mbox{for odd \textit{j}}).
\end{split}
\end{equation}

\subsection{Stationary solutions for the period-1 and period-2 states} 
\begin{figure}[b]
\includegraphics[height=2.7cm]{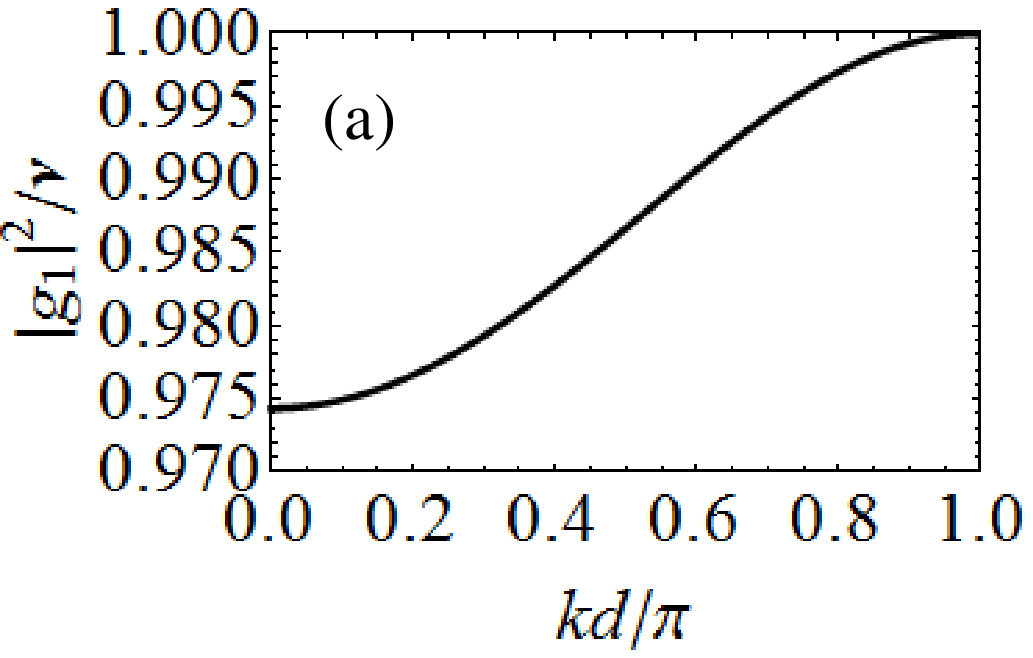}
\includegraphics[height=2.7cm]{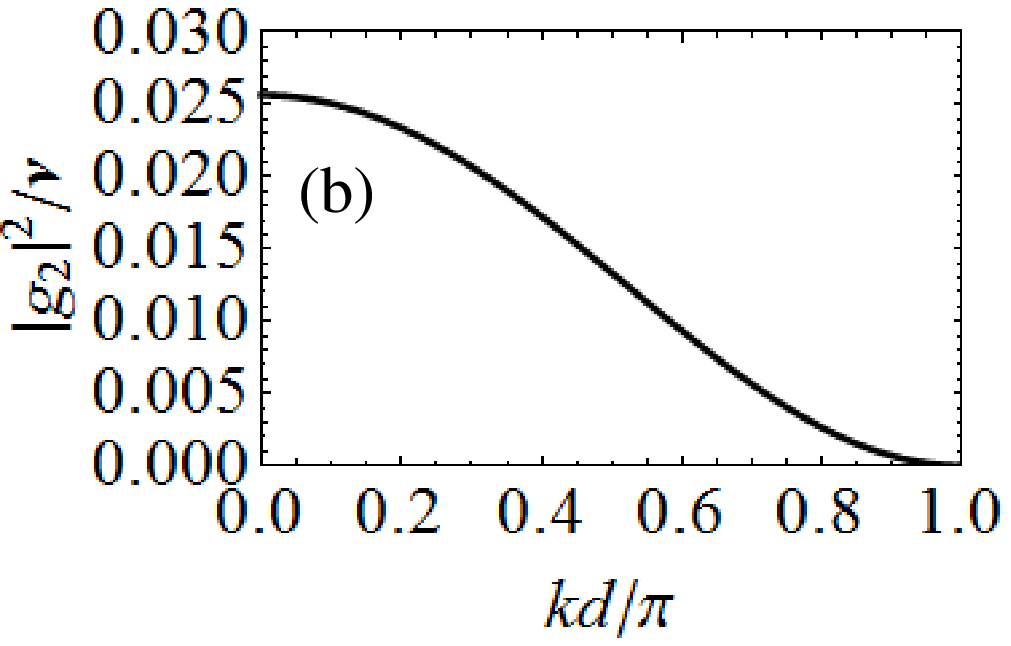}
\includegraphics[height=2.7cm]{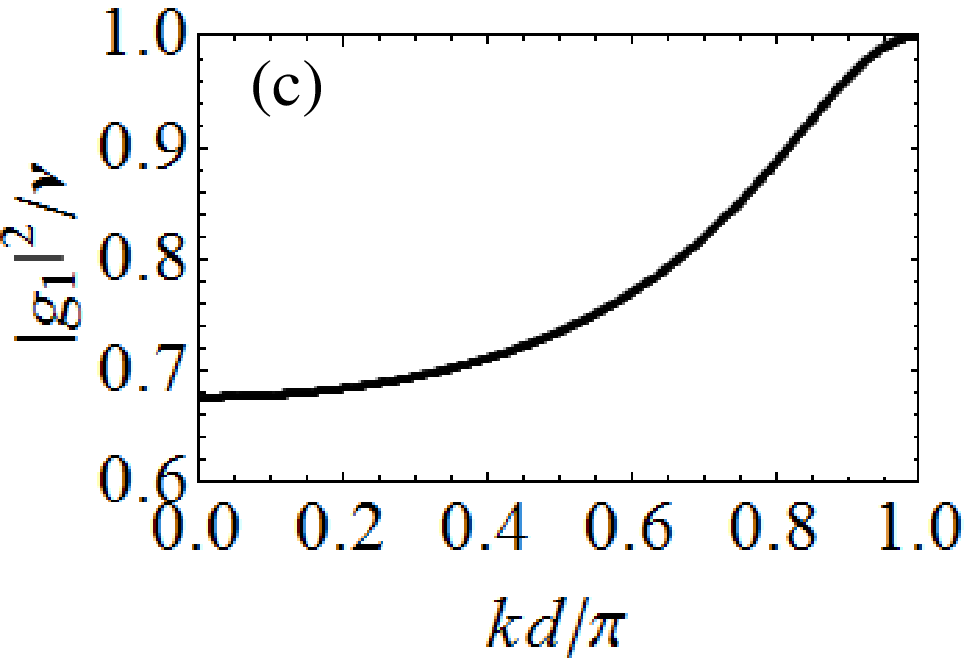}
\includegraphics[height=2.7cm]{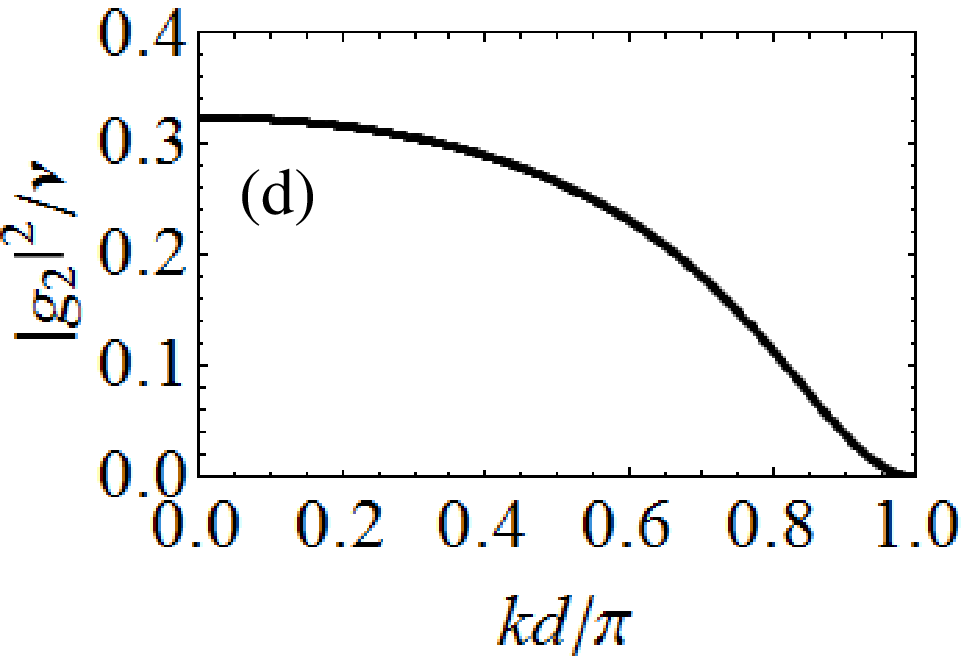}
\caption{Density distributions in the lowest band of the period-1 states as functions of $k$ for different values of $U\nu/2K$. Panels (a) and (b): $|g_1|^2$ (population in attractive site) and  $|g_2|^2$ (populations in repulsive site) for  $U\nu/2K=6$, respectively. Panels (c) and (d): $|g_1|^2$  and  $|g_2|^2$  for  $U\nu/2K=0.75$, respectively.}
\label{p1den}
\end{figure}

We focus on two particular cases: 1) period-1 states (normal Bloch states), i.e., when the particle density has the same periodicity as  that of the lattice, and 2) period-2 (period-doubled) states, i.e., when the particle density has twice the periodicity as  that of the lattice.
We separate from $\psi_j$ a plane-wave part, $e^{i k j \tilde{d}}$, and write $\psi_j$ in a product form: $g_j e^{i k j \tilde{d}}$, where $\hbar k$ is the quasimomentum of the bulk superflow flowing in the same direction of the lattice and $g_j$ is the complex amplitude at site $j$.

The period-1 unit cell consists of two lattice sites. Since the periodic boundary condition implies that $g_j=g_{j+2}$, we  have to solve Eq.~(\ref{var1}) for $g_1$ and $g_2$ only, subject to the condition 
\begin{equation}
|g_1|^2+|g_2|^2=\nu.
\end{equation}
Here, $\nu$ is the total number of particles in the unit cell with two sites.

The populations $|g_1|^2$ and $|g_2|^2$ in the attractive and the repulsive sites, respectively, for the lowest Bloch band are given by
\begin{equation}
  \frac{|g_1|^2}{\nu} = n_+ \quad\mbox{and}\quad \frac{|g_2|^2}{\nu} = n_-
\label{eq:pop}
\end{equation}
with
\begin{equation}
  n_{\pm} = \frac{1}{2} \left\{1 \pm \left[ \left(\frac{\cos{k\tilde{d}}}{U\nu/2K}\right)^2+1 \right]^{-1/2} \right\}\, .
\label{eq:npm}
\end{equation}

\begin{figure}[t]
\includegraphics[height=2.6cm]{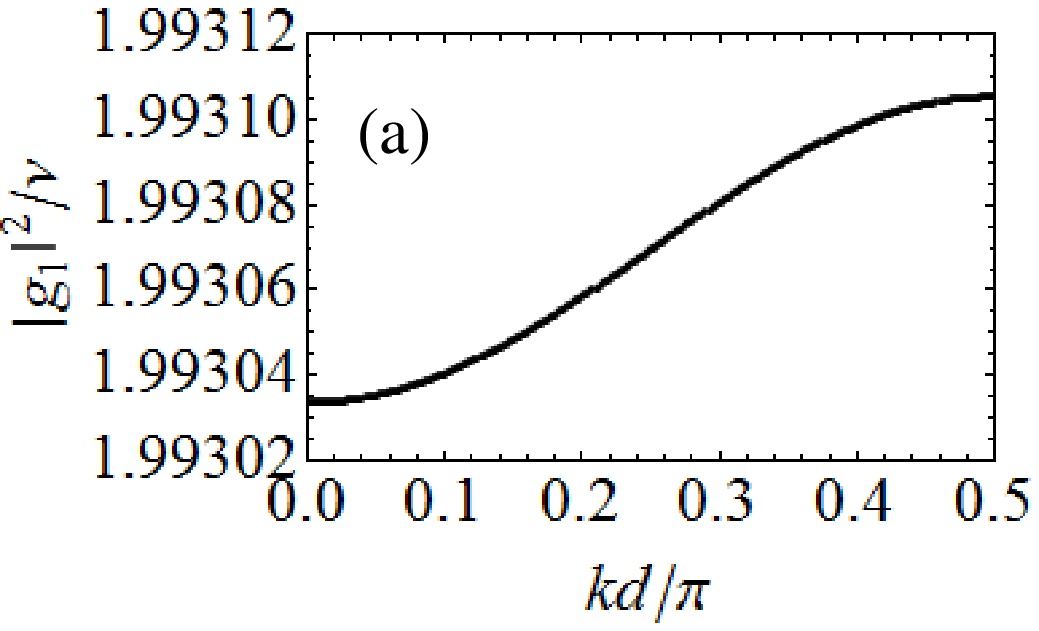}
\includegraphics[height=2.6cm]{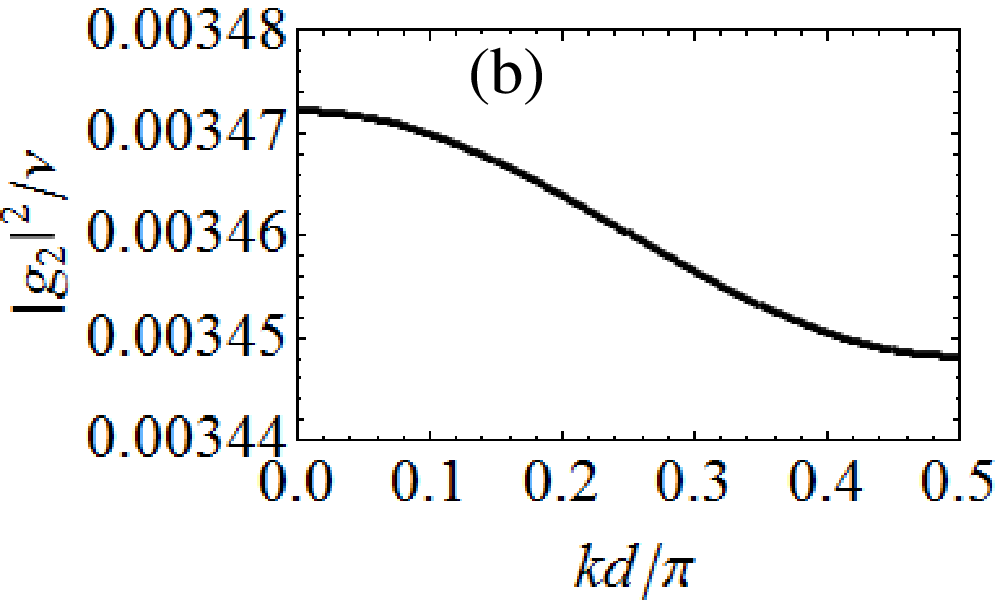}
\includegraphics[height=2.6cm]{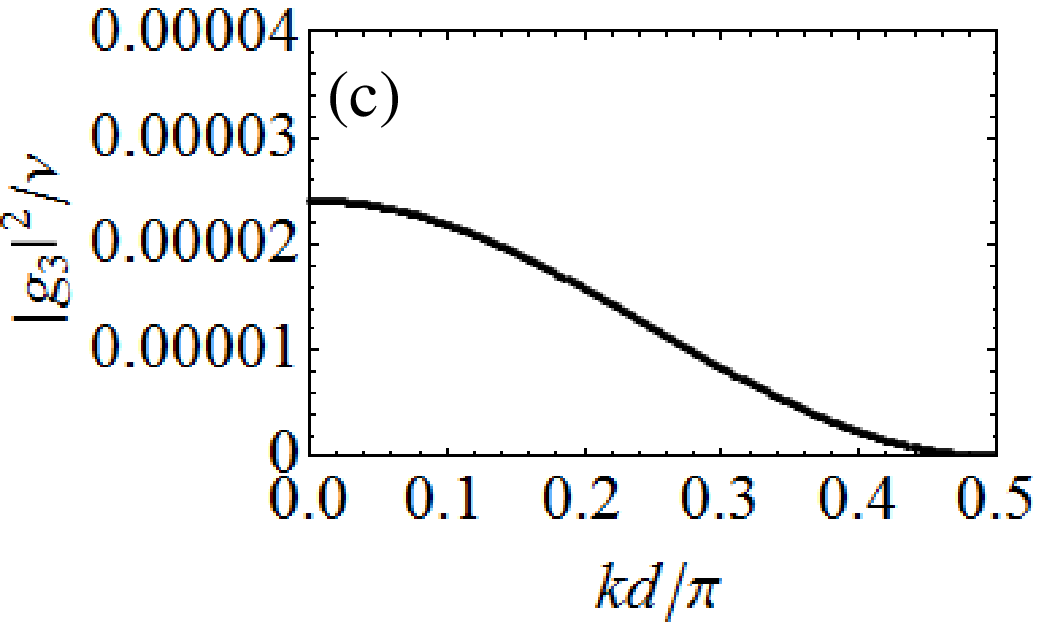}
\includegraphics[height=2.6cm]{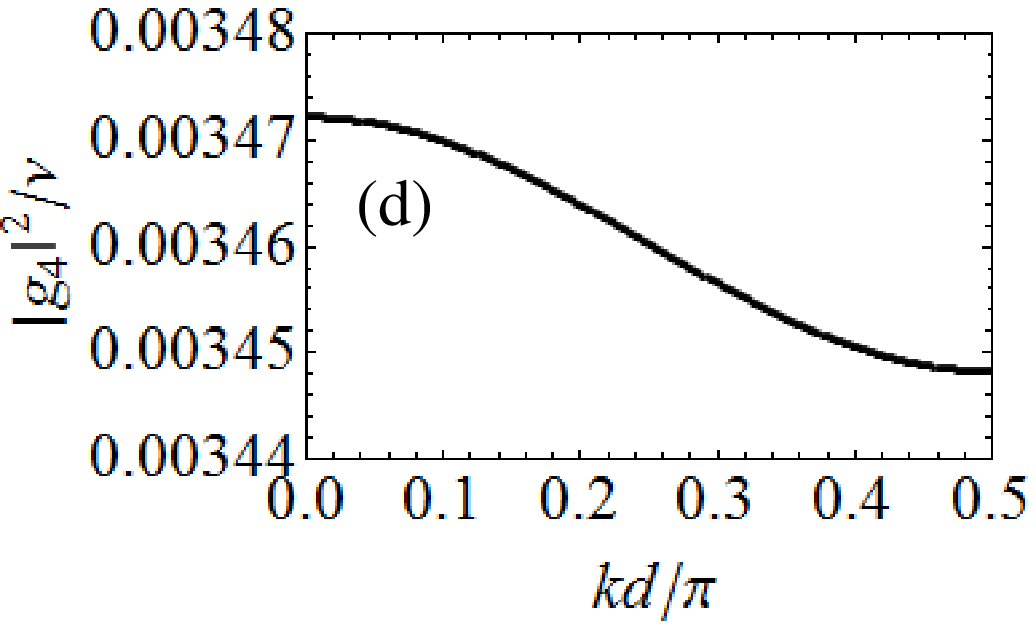}
\caption{Density distributions in the period-2 band for $U\nu/2K =6$: (a) $|g_1|^2$, (b) $|g_2|^2$, (c) $|g_3|^2$, and (d) $|g_4|^2$ (Populations in the 1st attractive site, 1st repulsive site, 2nd attractive site, and the 2nd repulsive site, respectively).}
\label{p2den1}
\end{figure}

\begin{figure}[b]
\includegraphics[height=2.7cm]{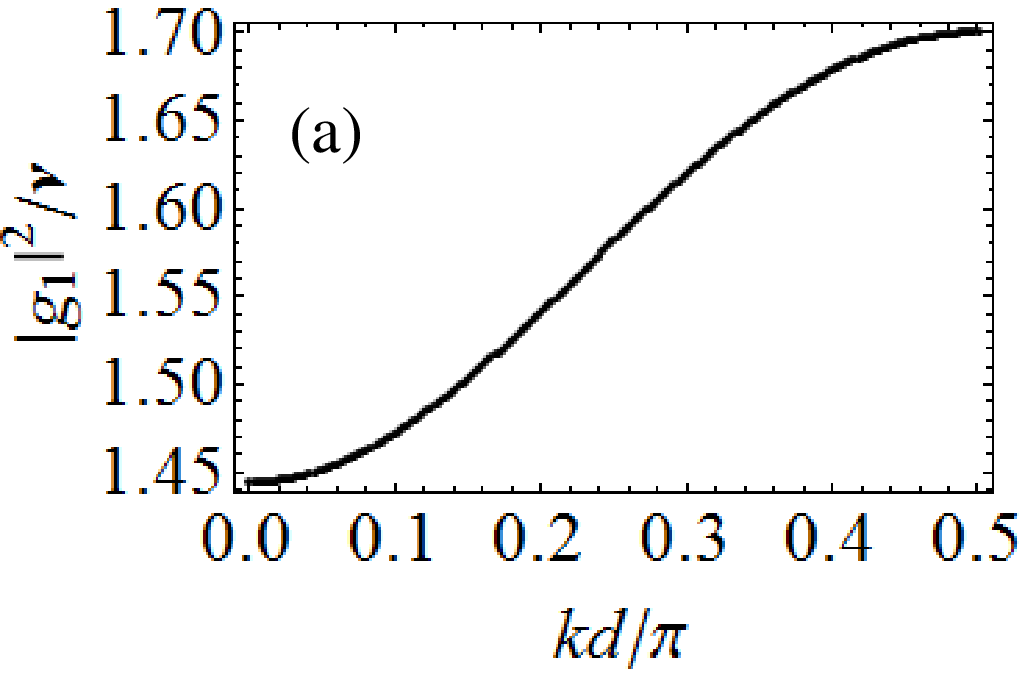}
\includegraphics[height=2.7cm]{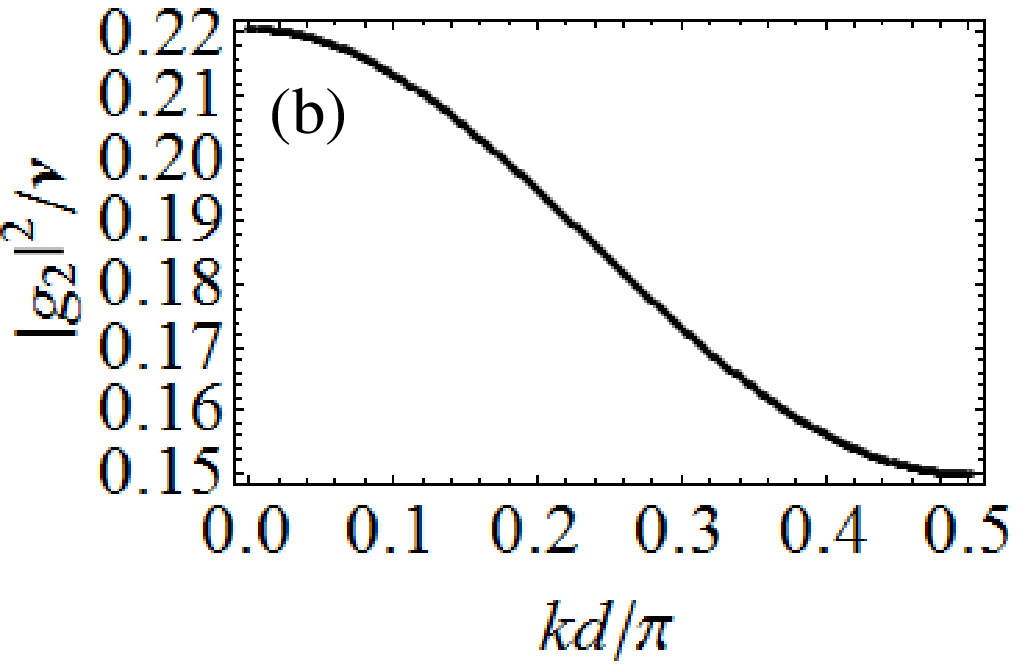}
\includegraphics[height=2.7cm]{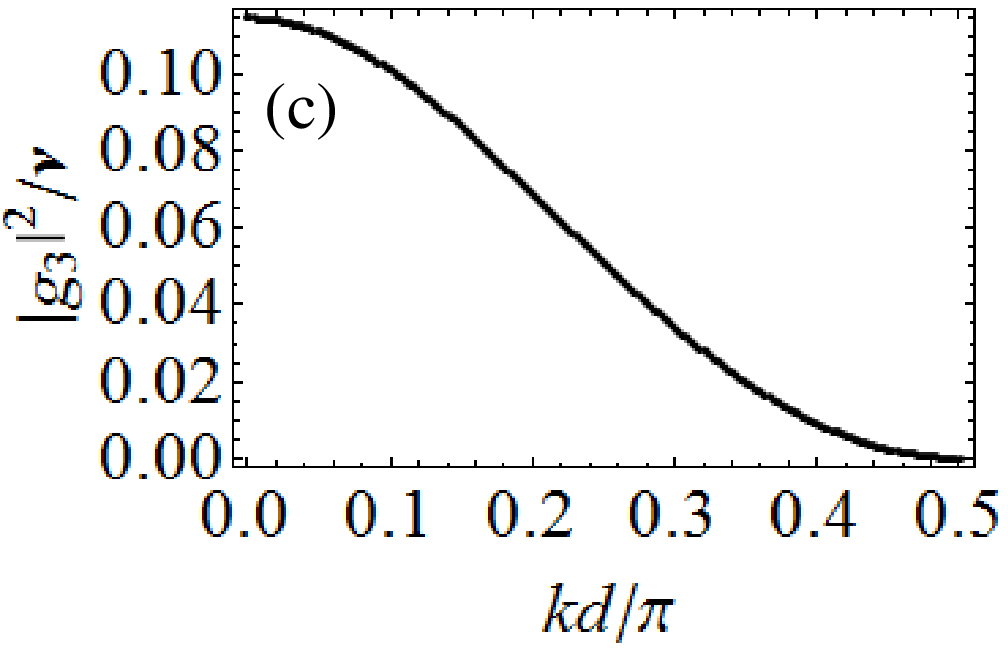}
\includegraphics[height=2.7cm]{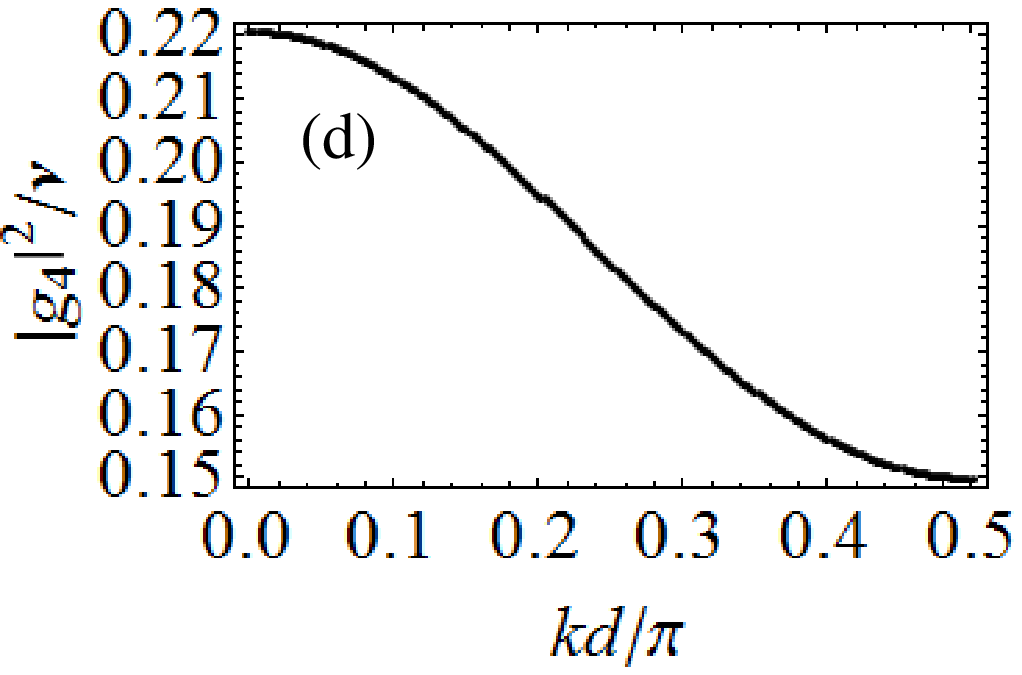}
\caption{The same as Fig.~\ref{p2den1} for $U\nu/2K =0.75$: (a) $|g_1|^2$, (b) $|g_2|^2$, (c) $|g_3|^2$, and (d) $|g_4|^2$.
}
\label{p2den2}
\end{figure}

The population density distributions for two different values of the dimensionless parameter $U\nu/2K$ are shown in Fig.~\ref{p1den} as functions of $k$ within the first Brillouin zone. We notice that when $U$ is sufficiently large [Figs. \ref{p1den}(a) and \ref{p1den}(b)], $|g_1|^2\approx\nu$ for all $k$ values. This can be easily understood from Eq.~(\ref{hamilt}): if $K\ll U$, putting all the particles in the attractive sites leads to the minimum-energy configuration of the system. In contrast, for smaller magnitudes of $U$, the kinetic-energy contribution also becomes significant. In this case, although at the zone edge most of the particles reside in the attractive sites, a sizable fraction of them is accumulated in the repulsive sites too, near the zone center [Figs. \ref{p1den}(c) and \ref{p1den}(d)]. 

For the period-2 case, the unit cell consists of four  lattice sites. The periodic boundary condition implies that $g_j=g_{j+4}$. So we  have to solve Eq. (\ref{var1})  for $g_1$, $g_2$, $g_3$, and $g_4$, subject to the condition 
\begin{equation}
|g_1|^2+|g_2|^2+|g_3|^2+|g_4|^2=2\nu.
\end{equation}
(Note that there is a factor of $2$ on the right-hand side since $\nu$ is defined as the number of particles per two-site unit cell.)

The distributions of $|g_1|^2$, $|g_2|^2$, $|g_3|^2$, and $|g_4|^2$ are shown for the period-doubled solutions with two different values of $U\nu/2K$ in Figs.~\ref{p2den1} and \ref{p2den2}. For a large $U\nu/2K$ (Fig.~\ref{p2den1}), the total energy is lowered by putting as many particles as possible in one attractive site in each supercell, i.e., in every fourth site. At the zone edge, the repulsive sites are almost empty and at the zone center they acquire a small population (Fig.~\ref{p2den1}). For a smaller $U\nu/2K$ (Fig.~\ref{p2den2}), the distribution is slightly more even: although one attractive site in a four-site cell hosts the majority of the particles, all the other sites, too, contain non-negligible populations.

\begin{figure}[t]
\includegraphics[scale=.4]{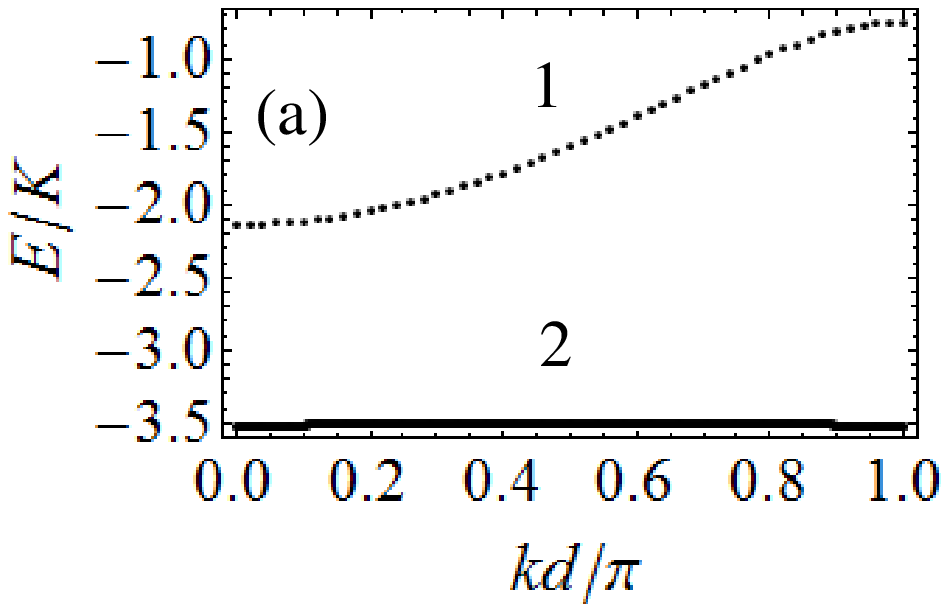}
\includegraphics[scale=.4]{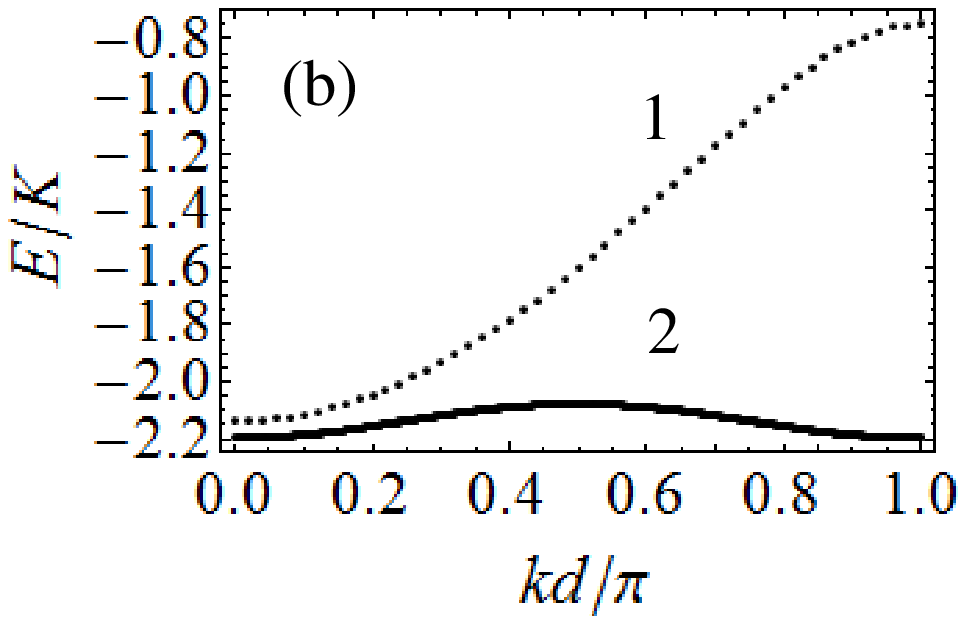}
\includegraphics[scale=.4]{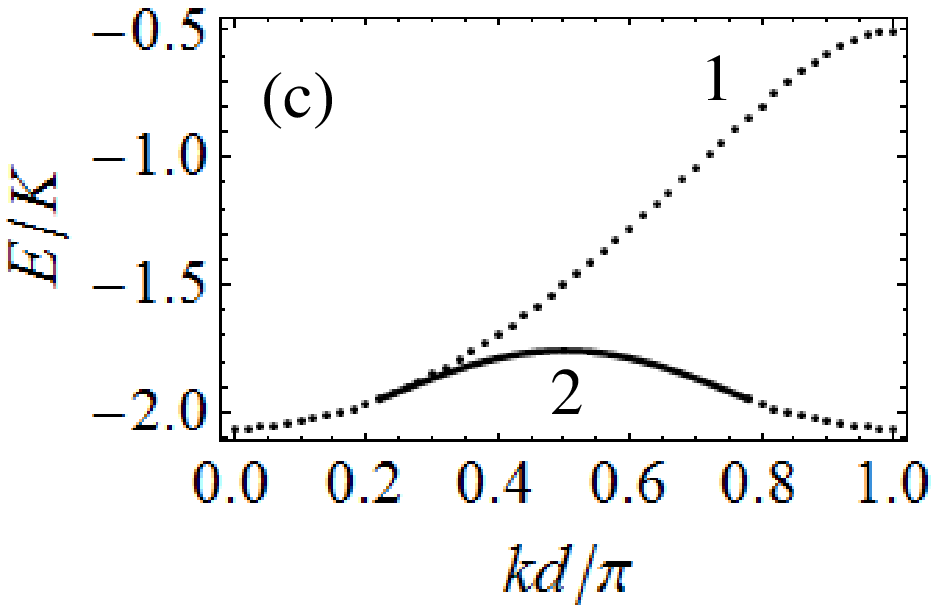}
\includegraphics[scale=.4]{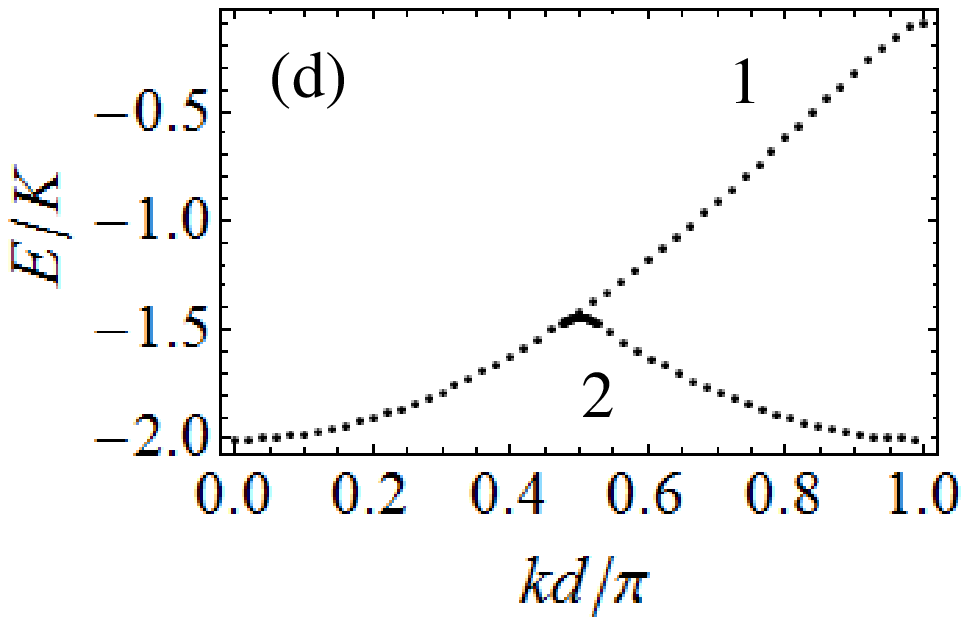}
\caption{Energy per particle of period-1 (dotted lines) and period-2 (solid lines) solutions in units of $K$ for different values of $U\nu/2K$: (a) $U\nu/2K=6$, (b) $U\nu/2K=0.75$, (c) $U\nu/2K=0.5$, and (d) $U\nu/2K=0.1$.}
\label{bands_d}
\end{figure}

Once we solve for the ${g_j}$'s, we can obtain the energy bands using Eq.~(\ref{hamilt}) with appropriate boundary conditions. The energy per particle, scaled by $K$ is a function of the dimensionless parameter $U\nu/2K$.  In Fig.~\ref{bands_d}, the period-1 (dotted line) and period-2 (solid line) bands are shown for four different values of $U\nu/2K$. We observe that when the nonlinear interaction term is large enough [Fig.~\ref{bands_d}(a)], the bands have a large separation between them and the period-2 band looks almost flat in comparison. For a relatively smaller value of $U\nu/2K$ [Fig.~\ref{bands_d}(b)], the gap between the two bands is narrower. Then if we keep lowering the value of $U\nu/2K$ [Fig.~\ref{bands_d}(c)], the two bands merge. In this case the period-2 band does not extend over the entire Brillouin zone, but appears in a small region centered around the zone edge, that shrinks further with decreasing $U\nu/2K$ [Fig.~\ref{bands_d}(d)]. 

For a given value of $U\nu/2K$, the period-2 bands show more flatness than their period-1 counterparts. As mentioned already, for period-2 states the majority of the particles are stored in every fourth site, while for period-1 states it is every second site. Thus, in the case of period-2 states, the degree of isolation between the regions of large density is higher.
This leads to a lower tunneling rate between consecutive sites. As a result, the energy bands are more flat for the period-2 case.

Also, a higher $U\nu/2K$ value leads to more relative flatness of the bands for both period-1 and period-2 solutions. This is because a large $U\nu/2K$ means that the on-site interaction term dominates over the hopping term and the stationary solutions are well approximated by the eigenstates of the on-site interaction term, which are independent of $k$. Another reason is that a large $U\nu/2K$ leads to repulsive sites being almost empty and the tunneling rate is suppressed.

\subsection{Linear stability analysis}

Let us now examine the stability of the stationary states of the system within the discrete model. There are two aspects: 1) energetic stability --- whether the stationary states are at a local energy minimum against small perturbations, and 2) dynamical stability --- if it is stable with respect to the time evolution.
As has been shown in general (see the Appendix of \cite{wu03}), energetic {\it instability} is a pre-requisite for dynamical {\it instability}. Namely, if the system is energetically stable, the system is dynamically stable as well; however, the opposite is not the case.

Here we perform a linear stability analysis of the stationary states following the treatment in Refs.~\cite{wu1,pethick1,wu2,pethick2} (see also, e.g., Refs.~\cite{pethickbook,wu03,nonlinlatrev}).
Let $\delta\psi_{q,j}$ be the deviation from the stationary solution $\psi^{(0)}_j$ at a given $k$,
\begin{equation}
\delta\psi_{q,j} = e^{ikj\tilde{d}} \left[ u_{q,j} e^{i q j \tilde{d}}+ {v_{q,j}}^*e^{-i q j \tilde{d}} \right], \label{eq:perturb}
\end{equation}
where the amplitudes $u_{q,j}$ and $v_{q,j}$ have the same periodicity as the stationary solution, $j$ is the site index, and $\hbar q$ is the quasimomentum of the perturbation. 
Now the energy functional in Eq.~(\ref{hamilt}) is expanded to second order in $\delta\psi_{q,j}$, and we find $\delta E_c$, its deviation from the equilibrium energy per unit cell. 

We can write $\delta E_c$ in a block-diagonal structure in $q$. For the period-1 case, it has the following form:
\begin{equation}
\delta E_c= \begin{pmatrix}
u_{q,1}^{*}& v_{q,1}^{*}& u_{q,2}^{*}& v_{q,2}^{*}
\end{pmatrix}
M(q)
\begin{pmatrix}
u_{q,1}\\
v_{q,1}\\
u_{q,2}\\
v_{q,2}\\
\end{pmatrix}\, .
\end{equation}
Because of the periodic boundary condition, we have $u_{q,j} =u_{q,j+2}$ and $v_{q,j}= v_{q,j+2}$. $M(q)$ is a $4\times4$ matrix, where
\begin{equation}
\begin{aligned}
\begin{split}
[M(q)]_{11}&=[M(q)]_{22}=U(|g_1|^2-|g_2|^2);\\
[M(q)]_{12}&=[M(q)]_{21}=U |g_1|^2;\\
[M(q)]_{13}&=[M^*(q)]_{31}=-Ke^{i(k+q) \tilde{d}};\\
[M(q)]_{24}&=-[M^*(q)]_{42}=-Ke^{-i(k-q) \tilde{d}};\\
[M(q)]_{33}&=[M(q)]_{44}=-U (|g_1|^2+|g_2|^2);\\
[M(q)]_{34}&=[M(q)]_{43}=U |g_2|^2\, ,
\end{split}
\end{aligned}
\end{equation}
and zero otherwise.

\begin{figure}[b]
\includegraphics[scale=.4]{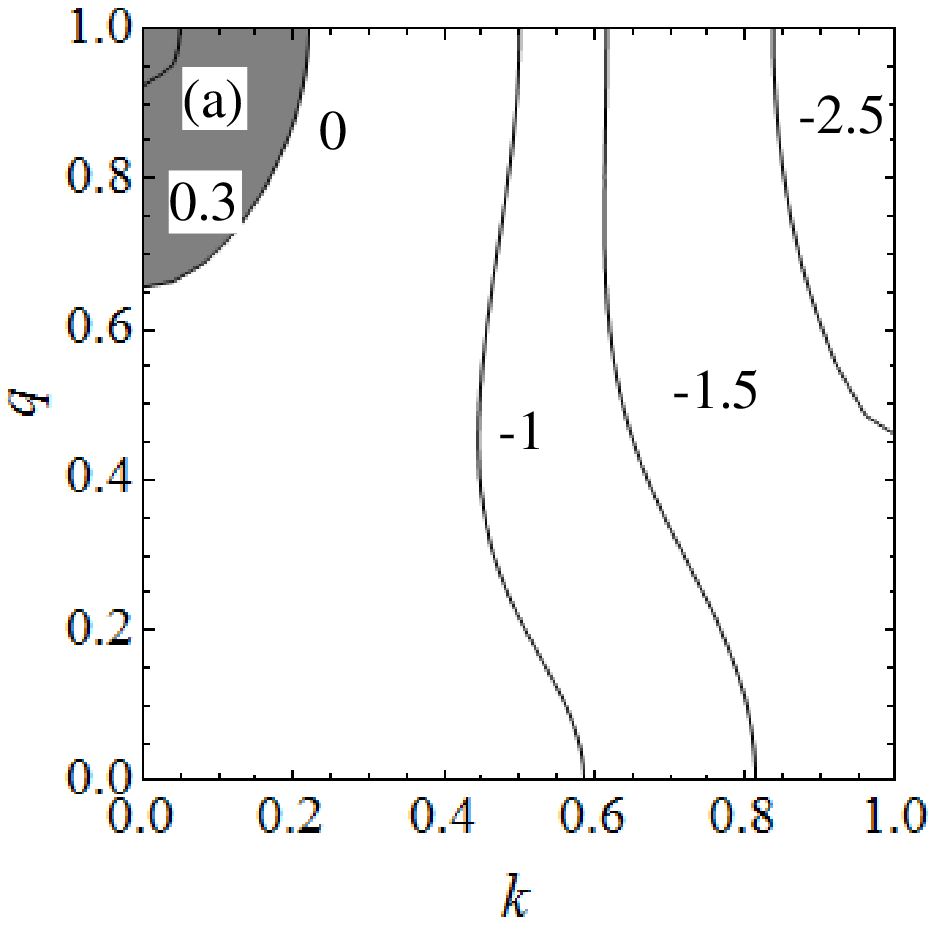}
\includegraphics[scale=.4]{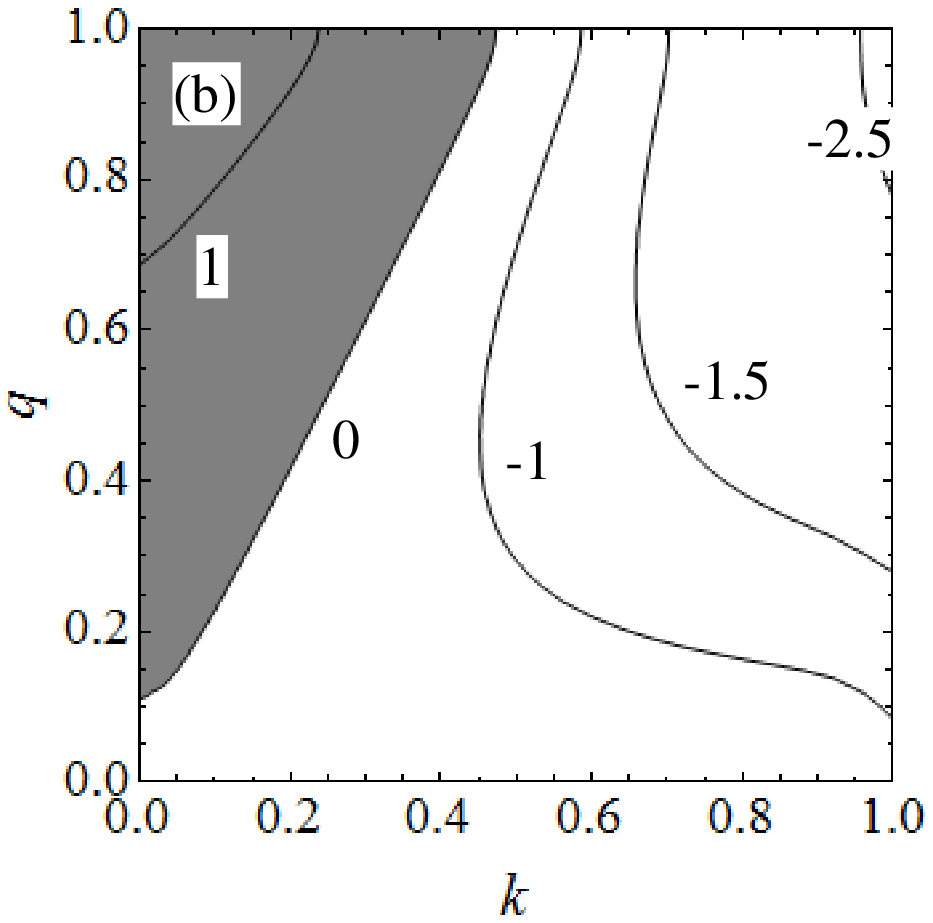}
\caption{Energetic stability diagrams for period-1 solutions for (a) $U\nu/2K=0.5$ and (b) $U\nu/2K=0.1$. Quasi-wave numbers $k$ and $q$ are in units of $k_0$. The gray-shaded regions are the energetically stable regions and the white regions are the energetically unstable regions. The contours show the minimum eigenvalue of the matrix $M(q)$ in units of $K$.
}
\label{d_enp1}
\end{figure}

\begin{figure}[b]
\includegraphics[scale=.4]{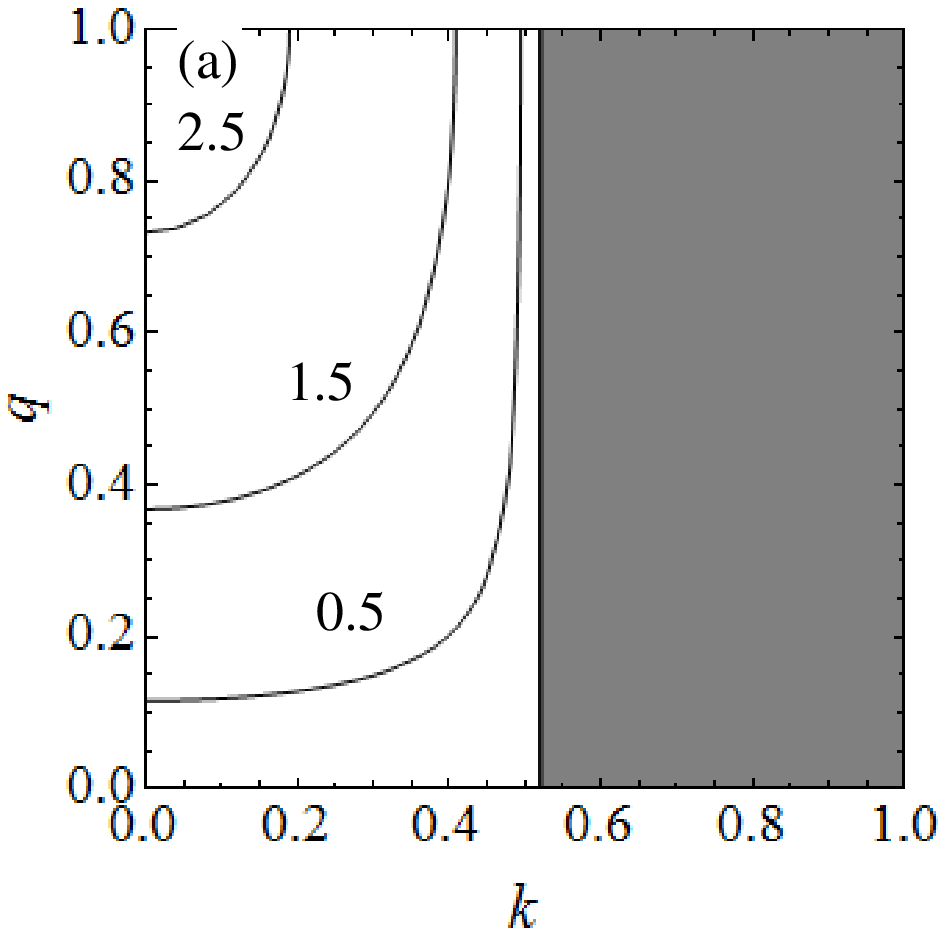}
\includegraphics[scale=.4]{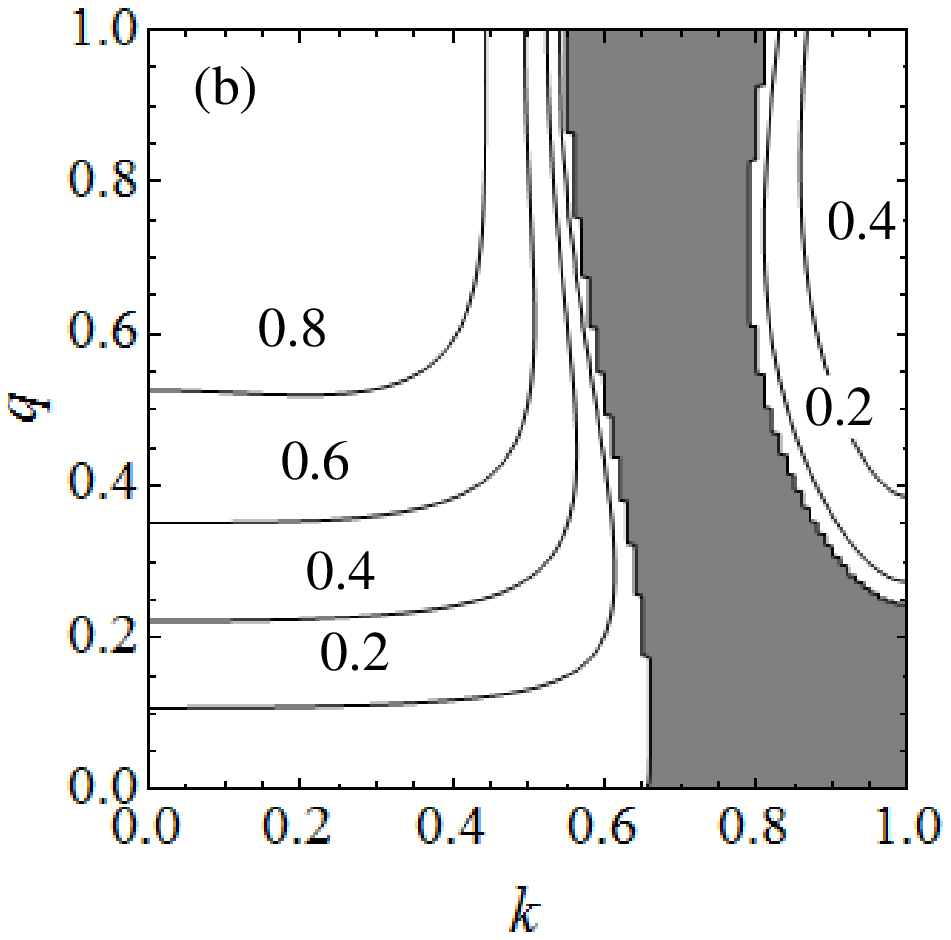}
\includegraphics[scale=.41]{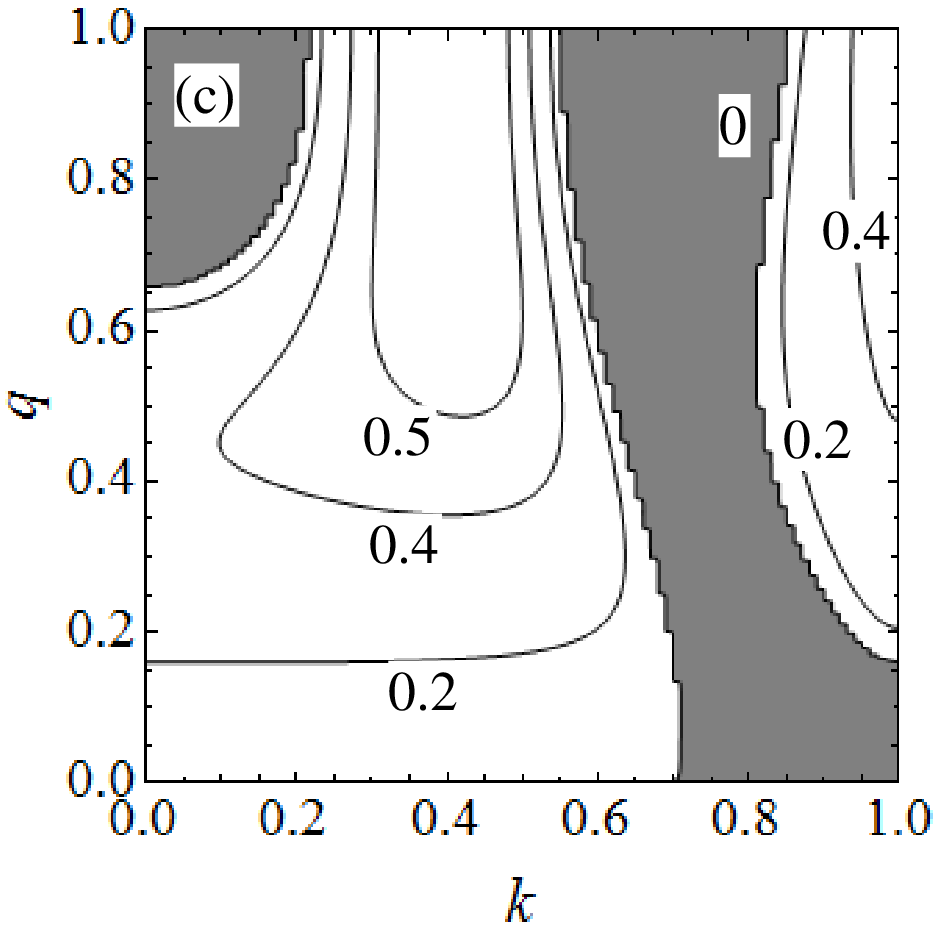}
\includegraphics[scale=.4]{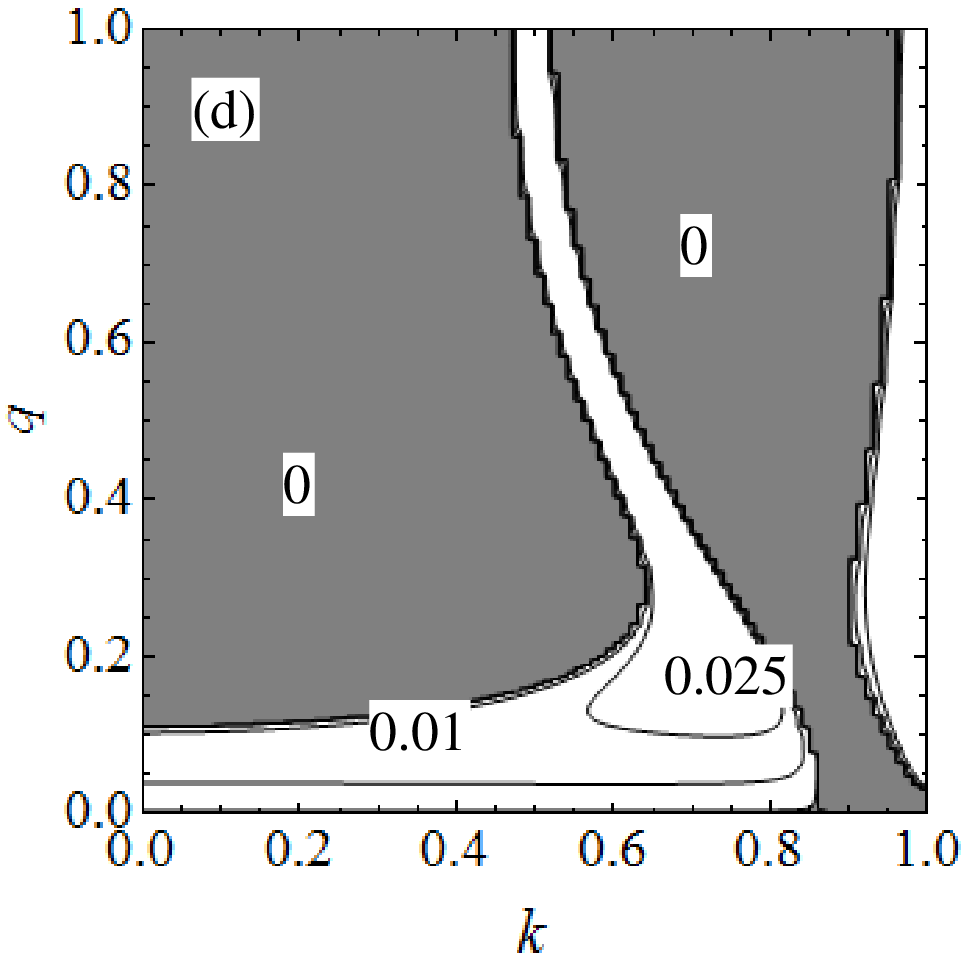}
\caption{Dynamical stability diagrams for period-1 solutions for different values of $U\nu/2K$: (a) $U\nu/2K=6$, (b) $U\nu/2K=0.75$, (c) $U\nu/2K=0.5$, and (d) $U\nu/2K=0.1$. Quasi-wave numbers $k$ and $q$ are in units of $k_0$.
The gray-shaded regions are the dynamically stable regions and the white regions are the dynamically unstable regions. The contours show the growth rate of the fastest growing mode, i.e., the maximum absolute value of the imaginary part of the eigenvalues of the matrix $M'(q)$ in units of $K$.
}
\label{d_dynp1}
\end{figure}

The condition for energetic stability of the system is that, all the eigenvalues of the matrix $M(q)$ are positive, since a negative eigenvalue means that there exist perturbations that can lower the energy of the system.  We thus study the energetic stability by noting the lowest eigenvalue of $M(q)$. If this value is $<0$, there exists at least one negative eigenvalue of $M(q)$, which would render the system energetically unstable. On the other hand, if this value is $\geqslant 0$, the system is already in either a local or global energy minimum, and hence stable.

We observe that for $U\nu/2K=6$ and $0.75$, no energetically stable region is found for period-1 solutions. An energetically stable area starts to appear for sufficiently low values of $U\nu/2K$ between $U\nu/2K=0.75$ and $0.5$ [see, e.g., $U\nu/2K=0.5$ and $0.1$ shown in Figs.~\ref{d_enp1}(a) and \ref{d_enp1}(b), respectively]. We show the instability contours, and the numbers on the lines mark the lowest eigenvalue of $M(q)$ for that parameter value. The stable regions are marked by the gray-shading.

We also consider the dynamical stability of the system under the same perturbation as Eq.~(\ref{eq:perturb}). The linearized time-dependent GP equation for the perturbations has the form
\begin{equation}
i \dfrac{\partial}{\partial t}\begin{pmatrix}
u_{q,1}\\
v_{q,1}\\
u_{q,2}\\
v_{q,2}\\
\end{pmatrix}=
M'(q)
\begin{pmatrix}
u_{q,1}\\
v_{q,1}\\
u_{q,2}\\
v_{q,2}\\
\end{pmatrix}\, .
\end{equation} 
Here $M'(q)$, too, is a $4\times 4$ matrix, where
\begin{equation}
M'(q)=\begin{pmatrix}
\sigma_z&0\\
0 & \sigma_z\\
\end{pmatrix}
M(q)
\end{equation}
with
\begin{equation}
\sigma_z =
  \begin{pmatrix}
    1 & 0\\
    0 & -1
  \end{pmatrix}.
\end{equation}

The condition for dynamical stability is that all the eigenvalues of the matrix $M'(q)$ are real, since a complex eigenvalue means that the perturbation grows exponentially in time during the dynamical evolution. We note the maximum of the absolute values of the imaginary parts of these eigenvalues to find out the fastest growing mode in the system. When this value happens to be zero, we get complete dynamical stability. 
 
The dynamical stability diagrams are shown in Fig.~\ref{d_dynp1}. It is found that the $k=0$ state is always unstable, so the superfluidity is not sustained in the Brillouin-zone center. This matches with the results obtained in \cite{wu1}, where they used the GP equation for the full continuum model to calculate the stationary states and study the corresponding stability properties. For higher values of $U\nu/2K$ [e.g., Fig.~\ref{d_dynp1}(a)], half the region between the Brillouin-zone center and the zone edge shows dynamical stability. If the value of $U\nu/2K$ is further reduced to $\sim 1$ [Fig.~\ref{d_dynp1}(b)], an instability island starts to grow from the zone edge. At even lower values of $U\nu/2K$, the instability region around the zone center starts to shrink [Fig.~\ref{d_dynp1}(c)], and we finally get a larger stability area [Fig.~\ref{d_dynp1}(d)]. Qualitatively, all these features are in agreement with the continuum-model results in \cite{wu1}.

We follow the same procedure for period-2 solutions to find the energetic and dynamic instabilities, only now both $M(q)$ and $M'(q)$ are $8\times 8$ matrices.  Moreover, for small values of $U\nu/2K$, the period-2 solutions do not exist for the entire Brillouin zone, but for a very small $k$-span near the zone edge.

\begin{figure}[h]
\includegraphics[scale=.4]{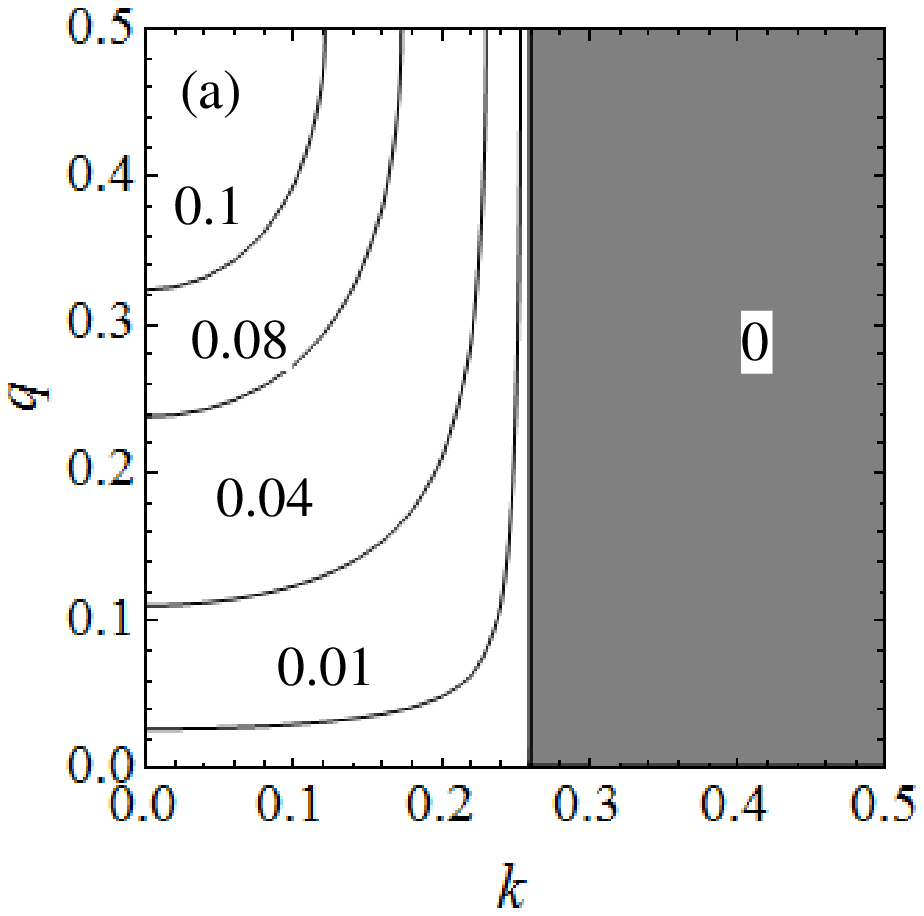}
\includegraphics[scale=.4]{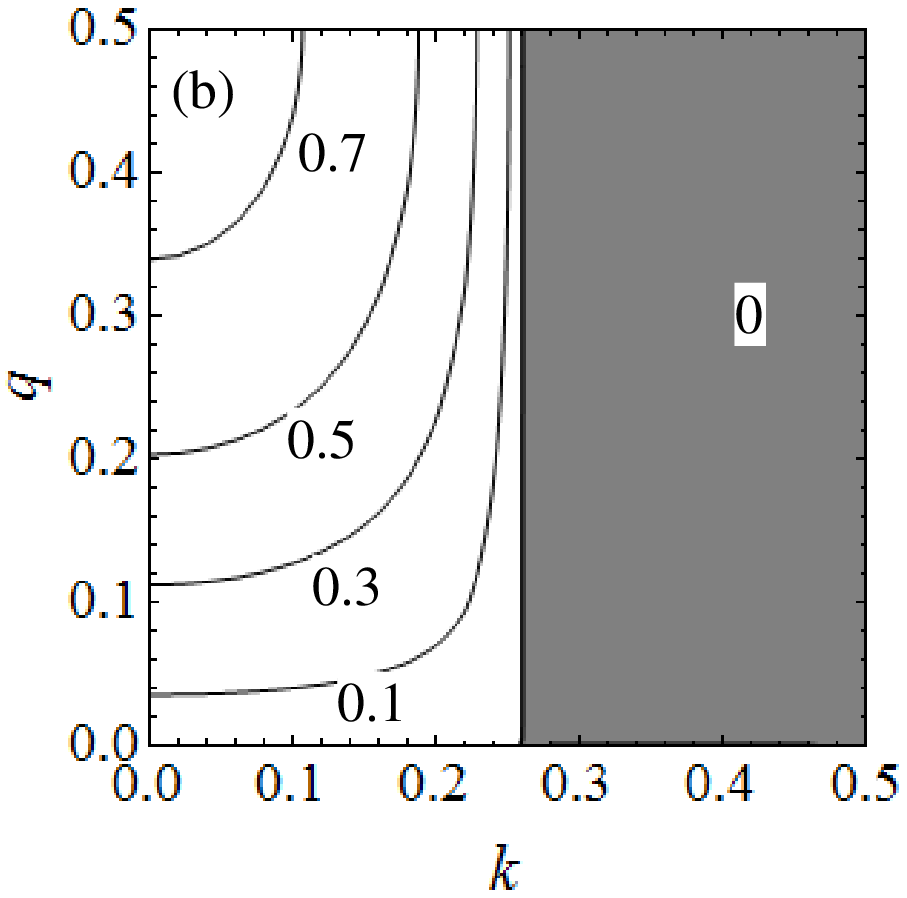}
\includegraphics[scale=.4]{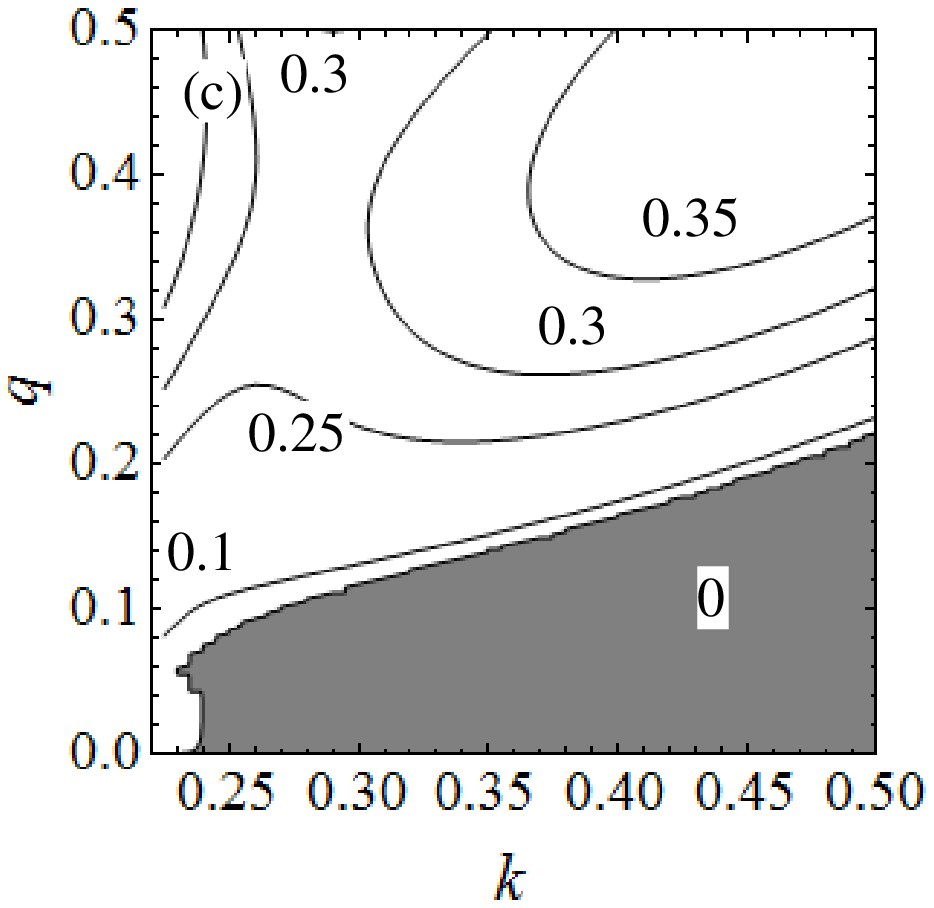}
\includegraphics[scale=.4]{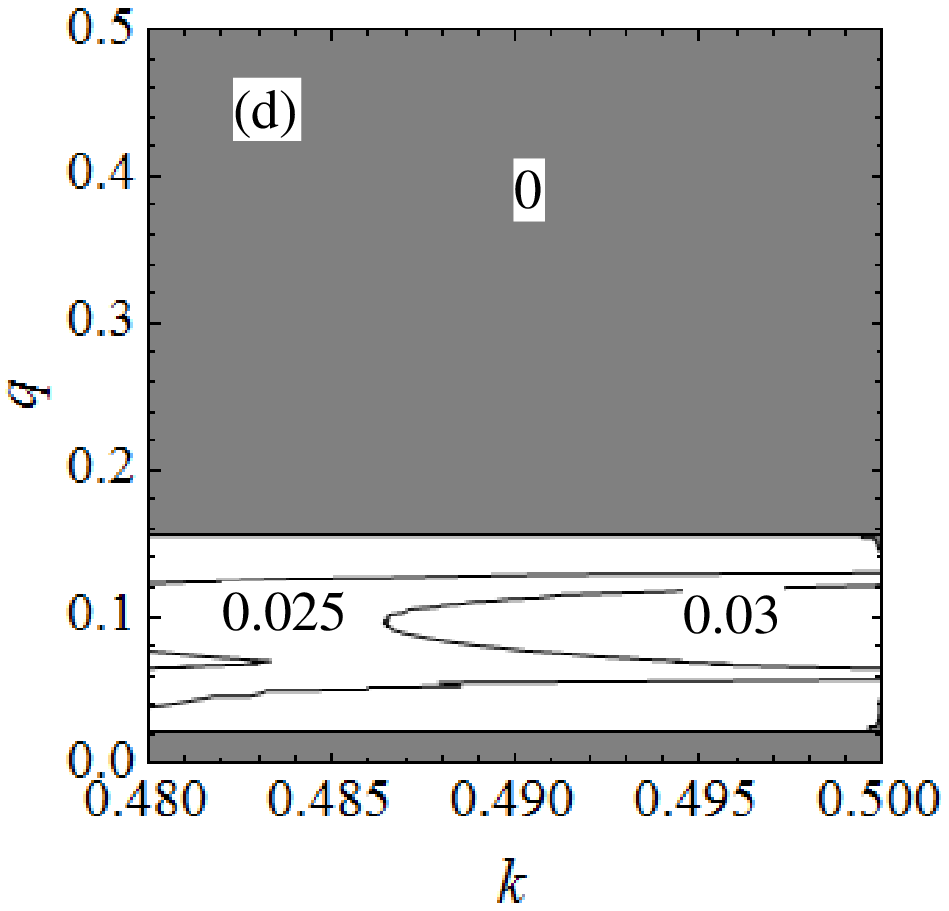}
\caption{The same as Fig.~\ref{d_dynp1} for period-2 solutions for (a) $U\nu/2K=6$, (b) $U\nu/2K=0.75$, (c) $U\nu/2K=0.5$, and (d) $U\nu/2K=0.1$.
}
\label{d_dynp2}
\end{figure}

As for the energetic stability, we now find that all the period-2 solutions are energetically unstable for the range of $U\nu/2K$ we are working with. For low $U\nu/2K$, the instability contours are horizontal. As $U\nu/2K$ is gradually increased, the contours become vertical, and the magnitude of the lowest eigenvalue of $M(q)$ (which is negative) becomes larger.

In the dynamical stability diagram for high nonlinearity [e.g., $U\nu/2K=6$ shown in Fig.~\ref{d_dynp2}(a)], the basic feature of the phase map remains the same as in the period-1 case. However, if we look at the contours of the fastest growing mode, here the value at $k=0$ is one order of magnitude smaller than the corresponding value for the period-1 case shown in Fig.~\ref{d_dynp1}(a) (25 times smaller if we consider high-$q$ perturbations). This point will be discussed in detail in Sec.~\ref{sec:mech}.

\section{The continuum model \label{sec:continuum}}
\subsection{Formalism and stationary solutions}

Next we turn to the continuum model, starting from the GP equation in 1D [Eq.~(\ref{GP1})]:
\begin{equation}
i \dfrac{\partial}{\partial t}\psi=-\dfrac{\partial^2}{\partial x^2}\psi+ (8 c_1+ 8 c_2 \mbox{cos}2x)|\psi|^2\psi\, .
\end{equation}
Here all the energies are measured in the scale of the recoil energy $E_R= \hbar^2 k_0^2/2 m$. All lengths are in units of $1/k_0$, and the time $t$ is in units of $2m/k_0^2\hbar$. The wave function $\psi$ is in units of $\sqrt{n_0}$, $n_0$ being the average number density. Here $c_1=n_0 V_1/8 E_R$ and $c_2=n_0 V_2/8 E_R$ (following the notation of \cite{wu1}). Again, we find solutions of the Bloch form, $\psi=e^{ikx}\phi$, where $\phi$ has the same periodicity as of the spatial modulation (period-1 solutions), twice the periodicity of it (period-2 solutions), or even higher period ones. To continue the analogy with the discrete model, we note that here, too, we can think of a ``supercell'', its length being $pd$ for a period-$p$ solution.

We expand $\psi$ in terms of plane waves,
\begin{equation}
\phi=\sum_{l= -l_{\rm max}}^{l_{\rm max}} a_l e^{i l x/p}
\end{equation}
($p$ is the periodicity of the solutions). Putting $p=1$ leads to the period-1 branches, while $p=2$ corresponds to period-doubled solutions.
Here $l$ can take $2l_{\rm max}+1$ values. The coefficients $a_l$ have to satisfy the normalization condition, $\sum_l |a_l|^2=1$. The stationary solutions are obtained by means of a variational calculation \cite{pethickbook}, so that the wave function $\psi(x)$ extremizes the total energy of the system.

\begin{figure}[t]
\includegraphics[scale=.4]{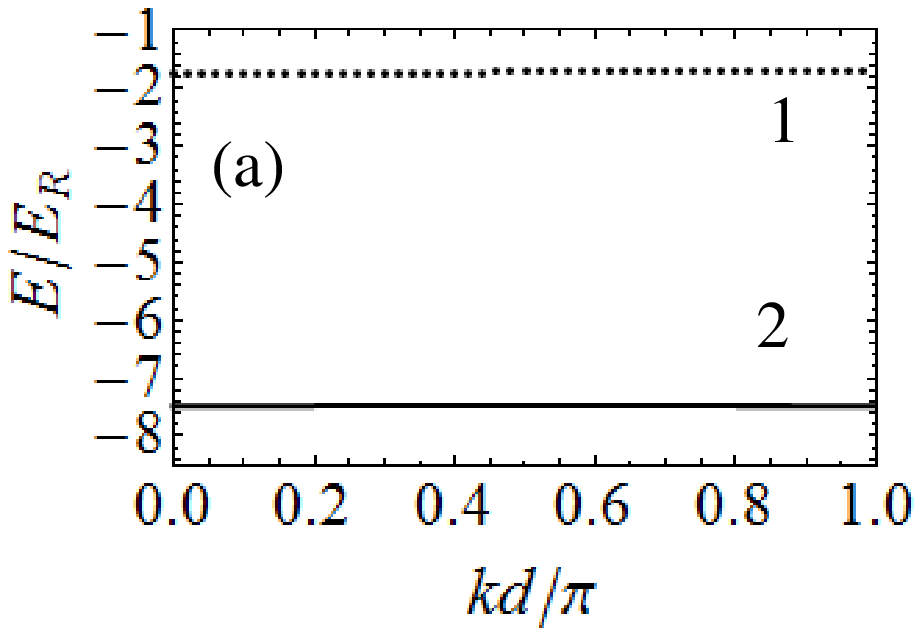}
\includegraphics[scale=.4]{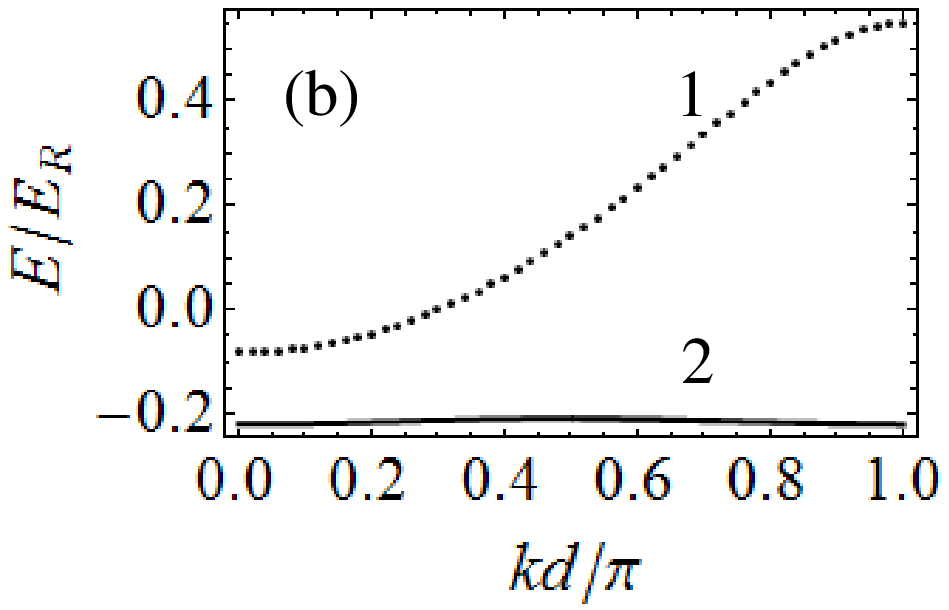}
\includegraphics[scale=.4]{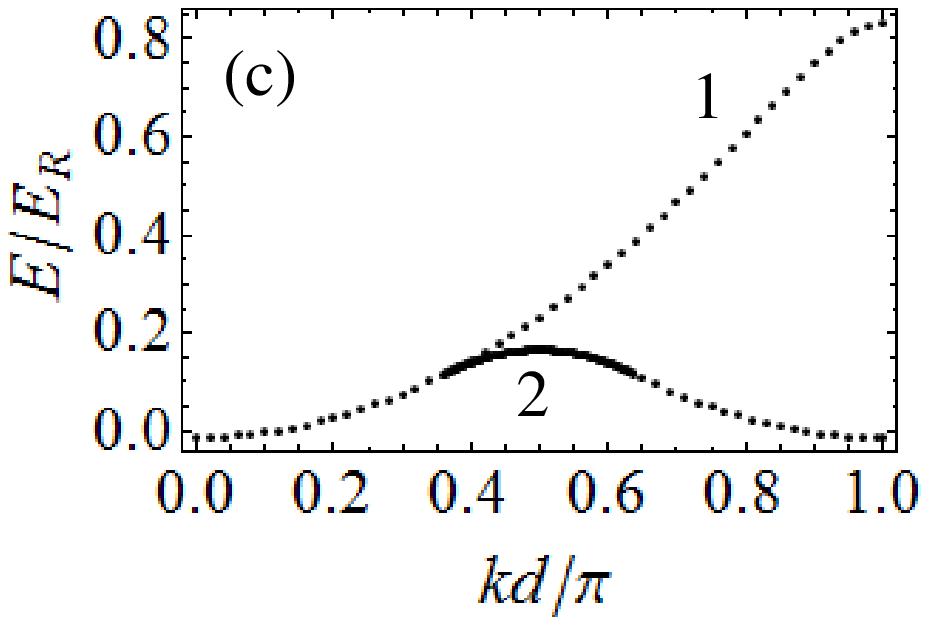}
\includegraphics[scale=.4]{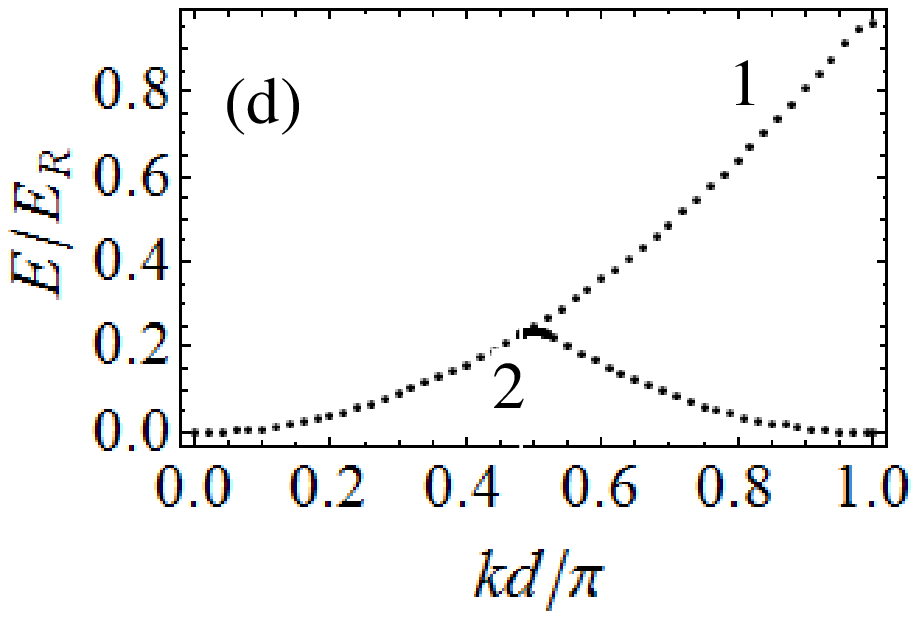}
\caption{Energy per particle of period-1 (dotted lines) and period-2 (solid lines) solutions for different values of $c_2$ obtained from the continuum model: (a) $c_2=0.4$, (b) $c_2=0.1$, (c) $c_2=0.04$, and (d) $c_2=0.01$.}
\label{band_cont}
\end{figure}

In Fig.~\ref{band_cont}, we show the energy bands corresponding to period-1 and period-2 solutions, for four different values of $c_2$, taking $c_1$=0. Just like the discrete case, we find  that when $c_2$ is large [Figs.~\ref{band_cont}(a) and \ref{band_cont}(b)], the bands are widely separated. As we keep decreasing the value of $c_2$ [Figs.~\ref{band_cont}(c) and \ref{band_cont}(d)], the two bands merge, and the region of the period-2 band starts diminishing. So our simplified discrete model can successfully capture all the essential features of the energy-band structures obtained from the full continuum calculation.

\begin{figure}[b]
\includegraphics[scale=.4]{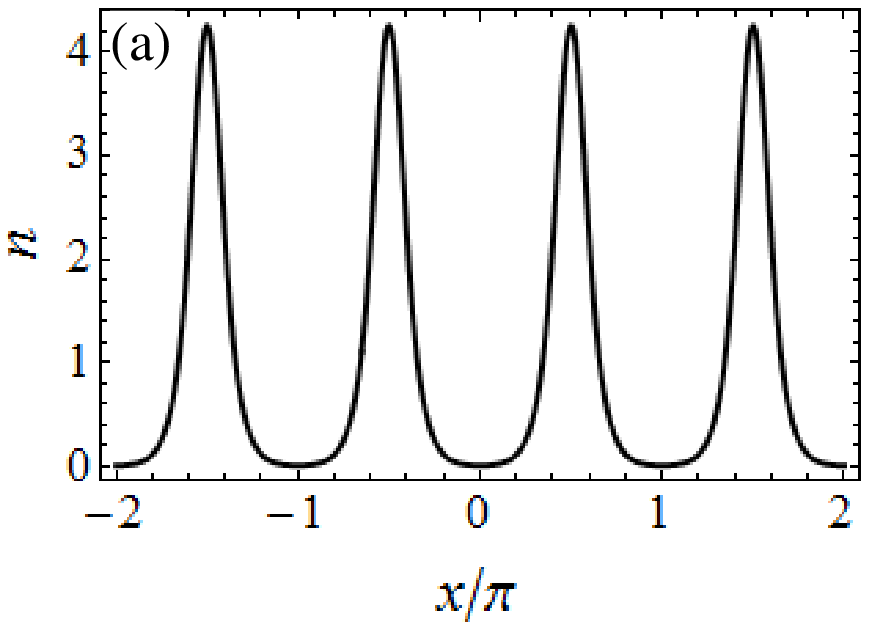}
\includegraphics[scale=.4]{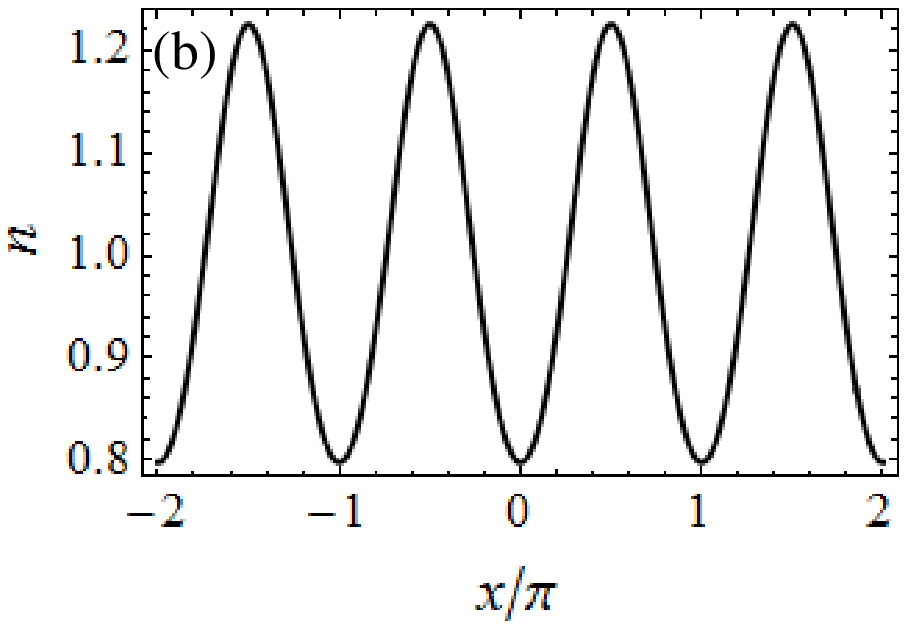}
\includegraphics[scale=.4]{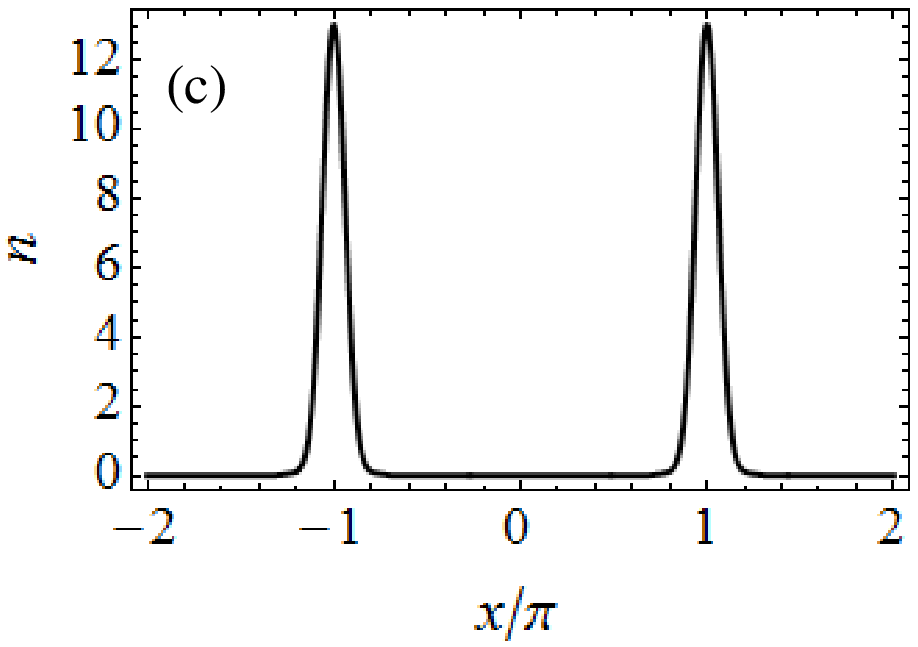}
\includegraphics[scale=.4]{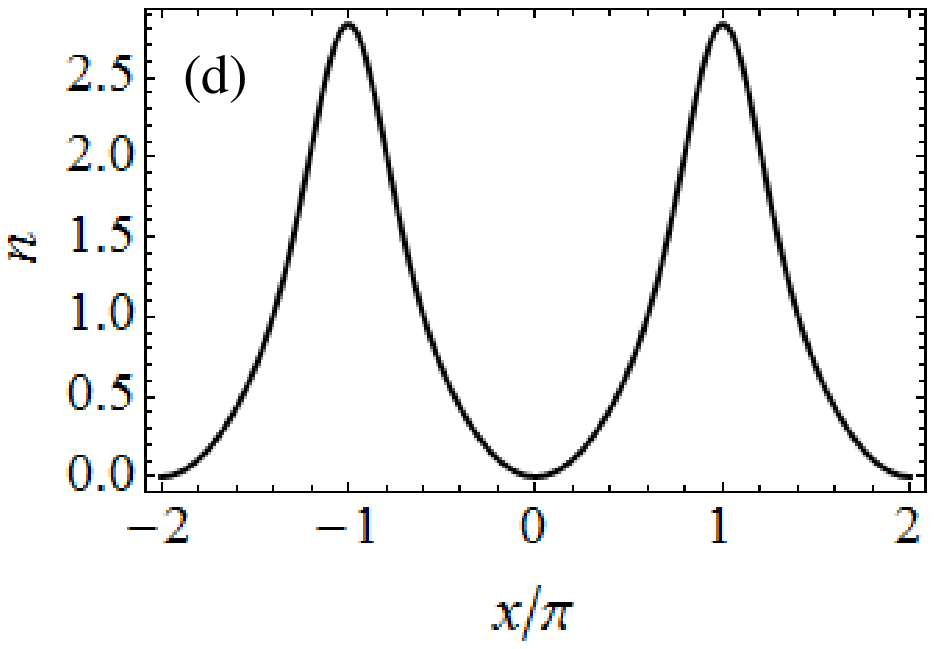}
\caption{Density distributions for (a) $c_2=0.4$, period-1, (b) $c_2=0.04$, period-1, (c) $c_2=0.4$, period-2, and (d) $c_2=0.04$, period-2, all for $k=0.5$ and $c_1=0$. Here $x$ is plotted in units of $1/k_0$, $n$ is in units of the average density $n_0$.
}
\label{c_den}
\end{figure}

Figure \ref{c_den} shows the nature of the density distribution in the continuum model, both for period-1 and period-2 solutions. It appears that for a fixed $c_1$, a larger $c_2$ makes the peaks sharper and more isolated in nature. 

We have chosen the $c_2$ values exactly as in \cite{wu1}, so that we can reproduce the stability diagrams from the period-1 case therein, before we proceed to solve for the period-2 case, and make a direct comparison. However, in this section we focus only on $c_1=0$ situations, because that corresponds to our discrete model of having alternate $U$ and $-U$ on-site interactions (a non-zero value of $c_1$ would mean that there is a difference in magnitude of the interaction strengths in the attractive and repulsive sites).

\begin{figure}[b]
\includegraphics[scale=.4]{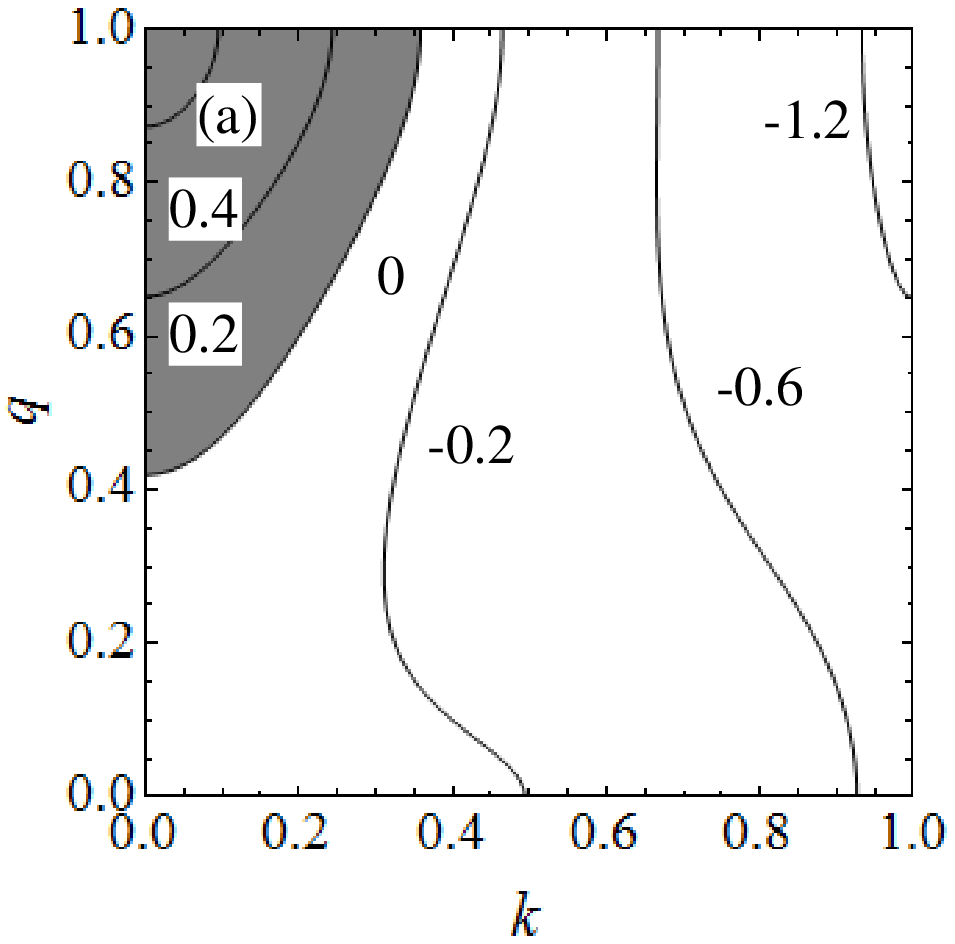}
\includegraphics[scale=.4]{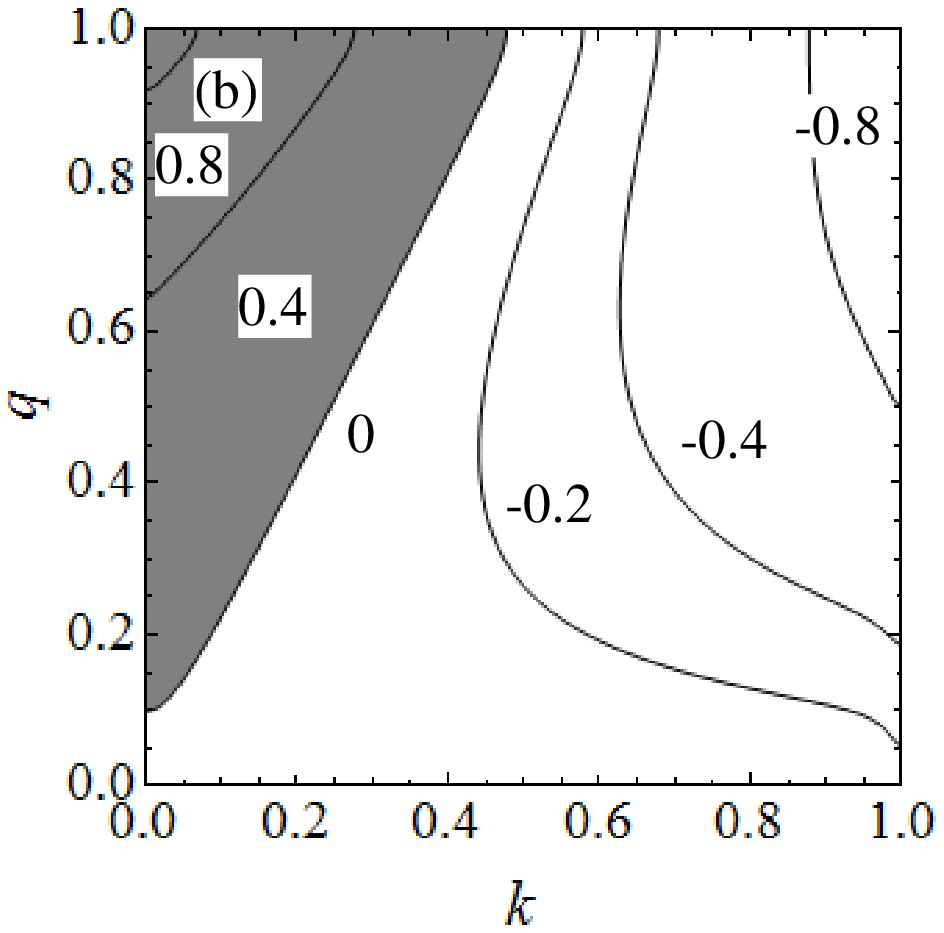}
\caption{Energetic stability diagrams for period-1 solutions for different values of $c_2$: (a) $c_2=0.04$ and (b) $c_2=0.01$. Quasi-wave numbers $k$ and $q$ are in units of $k_0$.
The gray-shaded regions are the energetically stable regions and the white regions are the energetically unstable regions. The contours show the minimum eigenvalue of the matrix $M(q)$ in units of the recoil energy $E_R$.
}
\label{en_contp1}
\end{figure}

\subsection{Stability analysis for the continuum model}

Let $\delta\psi_{q}$ be the deviation from the stationary Bloch wave solution $\psi^{(0)}$ at a given $k$ for the continuum model. This can be written as 
\begin{equation}
\delta\psi_{q}(x) = e^{ikx} \left[ u(x,q) e^{i q x}+ {v}^*(x,q)e^{-i q x} \right],
\end{equation}
where the amplitudes $u(x,q)$ and $v(x,q)$ are periodic functions of $x$ with the same periodicity as the stationary solutions. Similarly to the discrete model, the energy deviation from the stationary states per unit cell is given by
\begin{equation}
\delta E_c=\int_{-p \pi/2}^{p \pi/2} dx \begin{pmatrix}
u^{*}& v^{*}
\end{pmatrix}
M(q)
\begin{pmatrix}
u\\
v\\
\end{pmatrix}
\end{equation}
for $p$-periodic states.

We proceed exactly like in the case of the discrete model, and find the eigenvalues for $M(q)$, both for period-1 and period-2 solutions. If $M(q)$ has negative eigenvalues, that would render the system energetically unstable. In the period-1 case, a higher value of $c_2$ makes the system completely unstable energetically, while for smaller $c_2$ an energetically stable region (marked by the gray shade in Fig.~\ref{en_contp1}) appears, as in \cite{wu1}. For period-2 cases, the solutions are always unstable energetically, at least for the range of $c_2$ we have chosen, namely $0.01 \leq c_2 \leq 0.4$. This is exactly in agreement with the result we obtained in the discrete model.

The dynamical stability for period-1 and period-2 solutions is also studied. For the same perturbation $\delta\psi_q$, the time-dependent GP equation can be linearized as
\begin{equation}
i \dfrac{\partial}{\partial t}\begin{pmatrix}
u\\
v\\
\end{pmatrix}=
M'(q)
\begin{pmatrix}
u\\
v\\
\end{pmatrix}\,
\end{equation}
with $M'(q) \equiv \sigma_z M(q)$.
If $M'(q)$ has complex eigenvalues, the perturbations blow up in the course of time evolution, and if the imaginary part is zero, the stationary solutions are dynamically stable. The fastest growing modes (the mode with the largest absolute value of the imaginary parts of the eigenvalues) in the system are also noted.

\begin{figure}[t!]
\includegraphics[scale=.4]{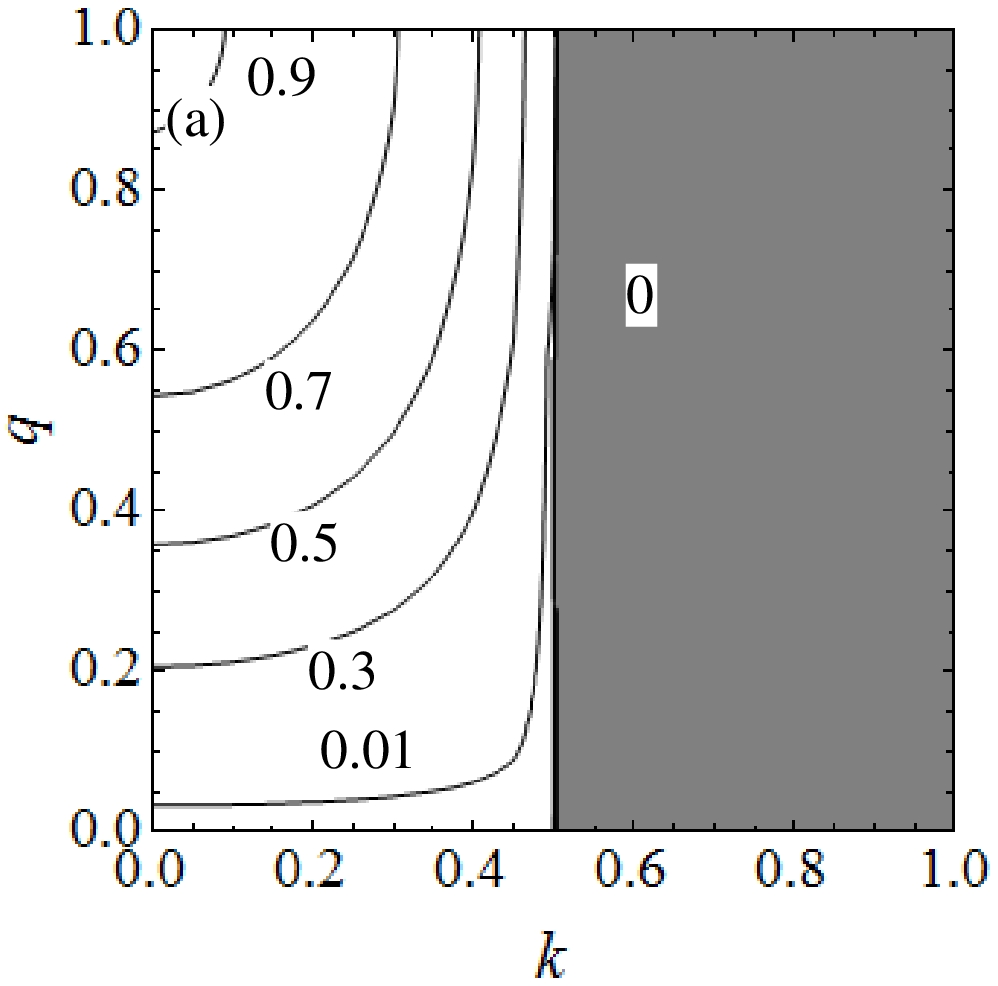}
\includegraphics[scale=.4]{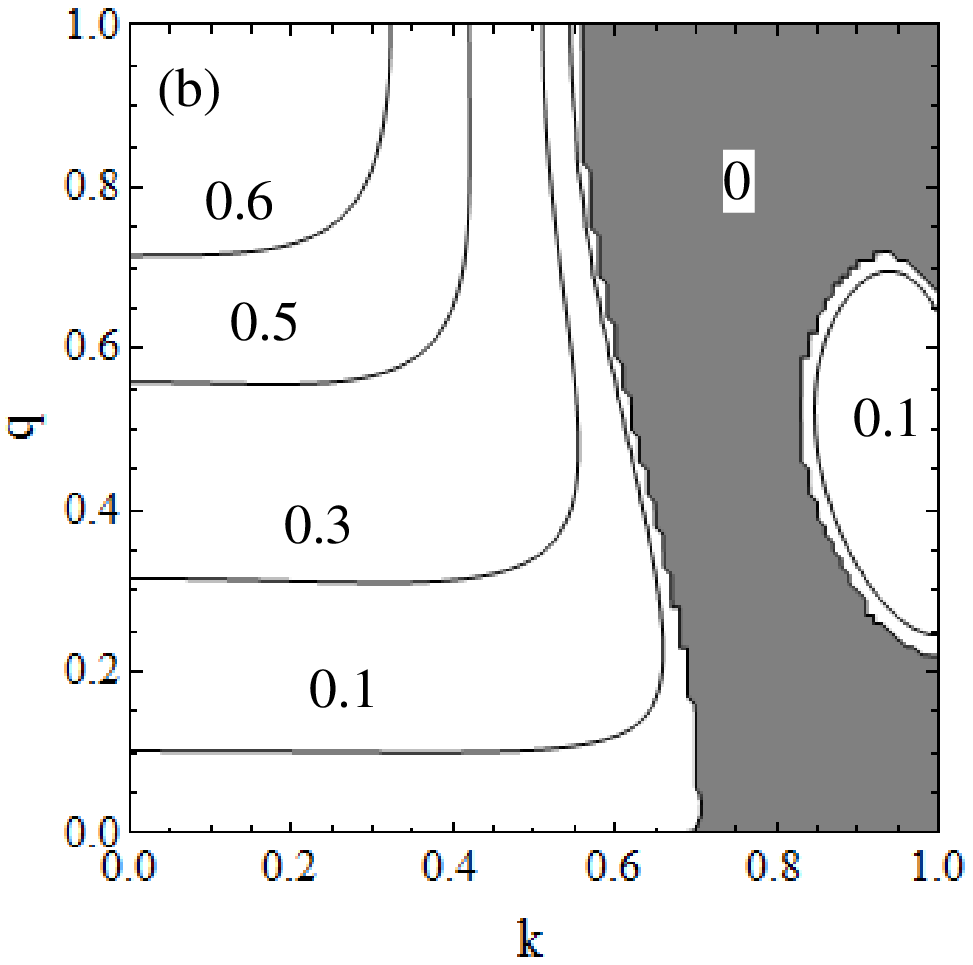}
\includegraphics[scale=.4]{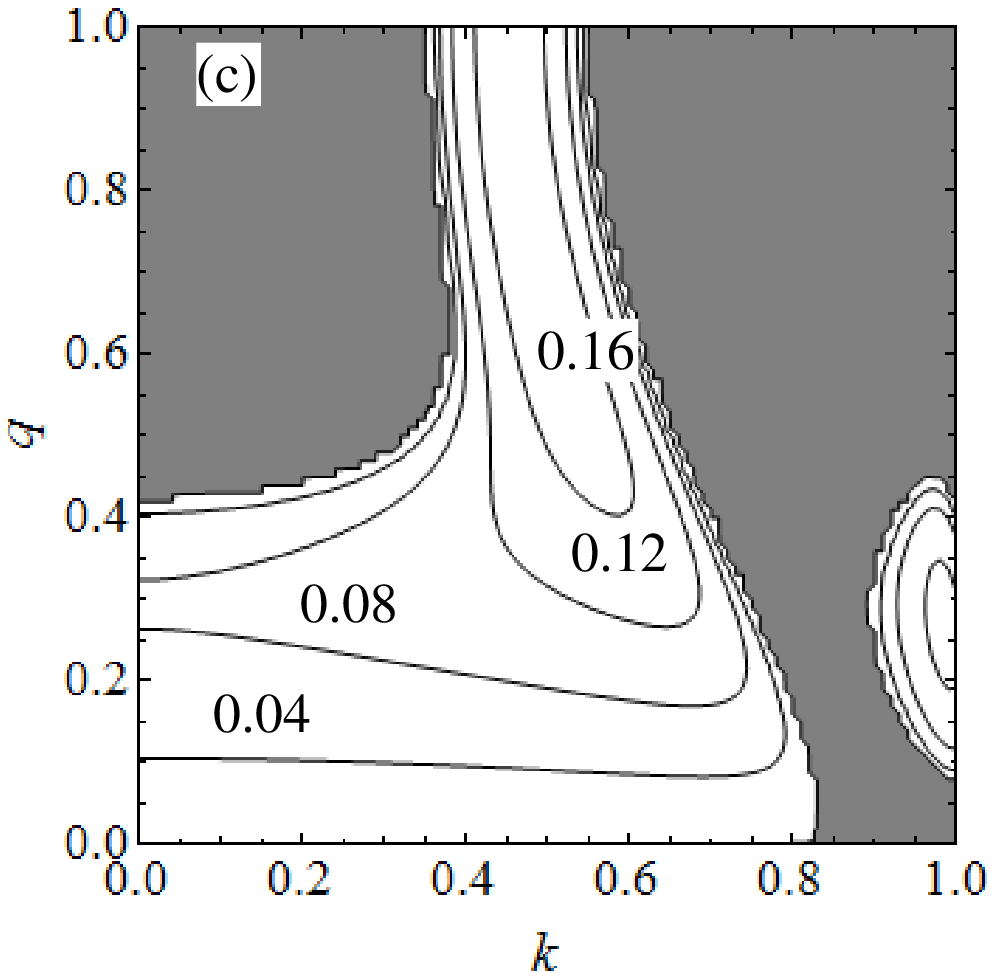}
\includegraphics[scale=.4]{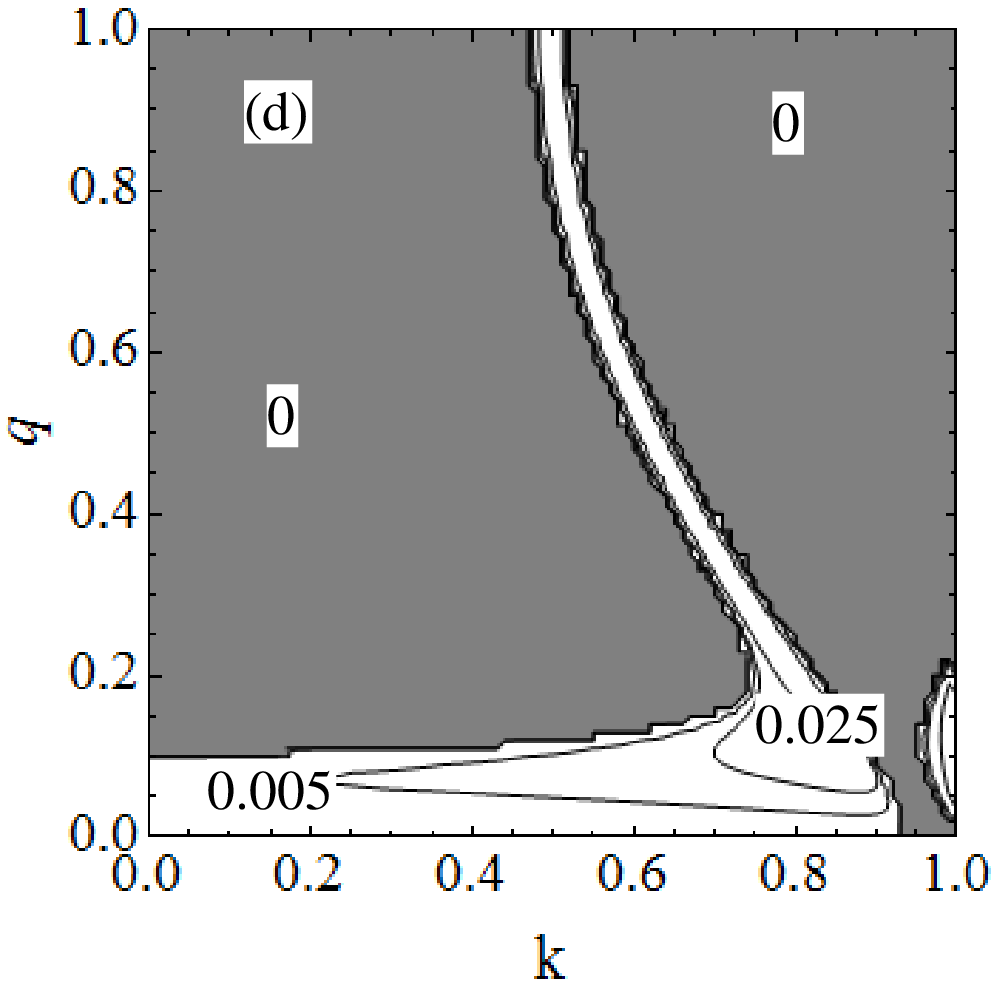}
\caption{Dynamical stability diagrams for period-1 solutions for different values of $c_2$: (a) $c_2=0.4$, (b) $c_2=0.1$, (c)$c_2=0.04$, and (d) $c_2=0.01$. Quasi-wave numbers $k$ and $q$ are in units of $k_0$.
The contours show the growth rate of the fastest growing mode, i.e., the maximum absolute value of the imaginary part of the eigenvalues of the matrix $M'(q)$, in units of the recoil energy $E_R$.
}
\label{c_dynp1}
\end{figure}
\begin{figure}[t]
\includegraphics[scale=.4]{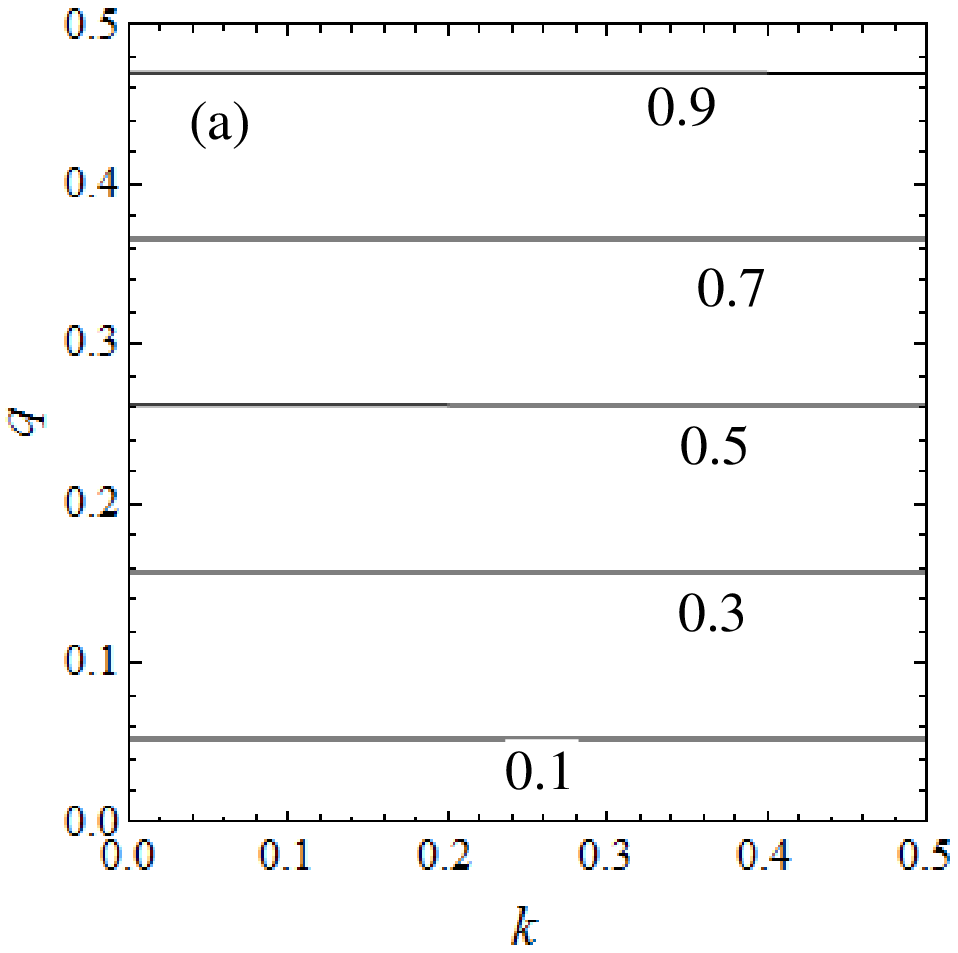}
\includegraphics[scale=.4]{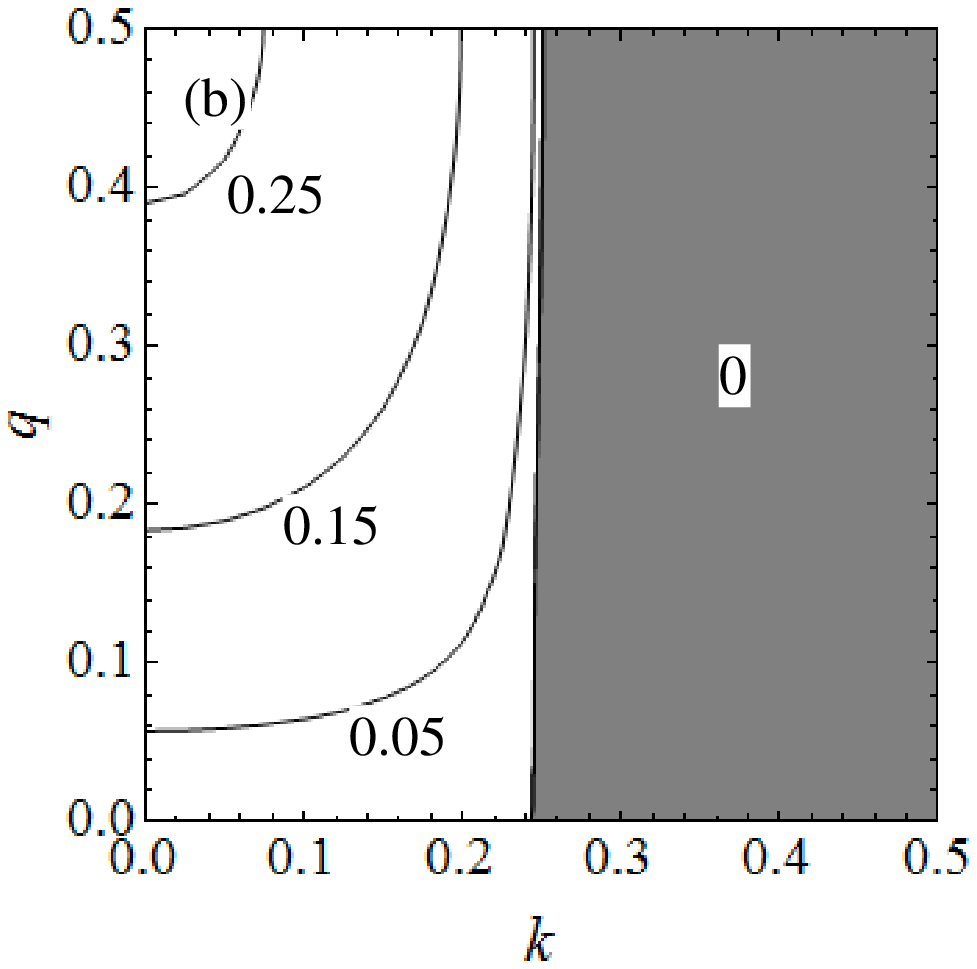}
\includegraphics[scale=.4]{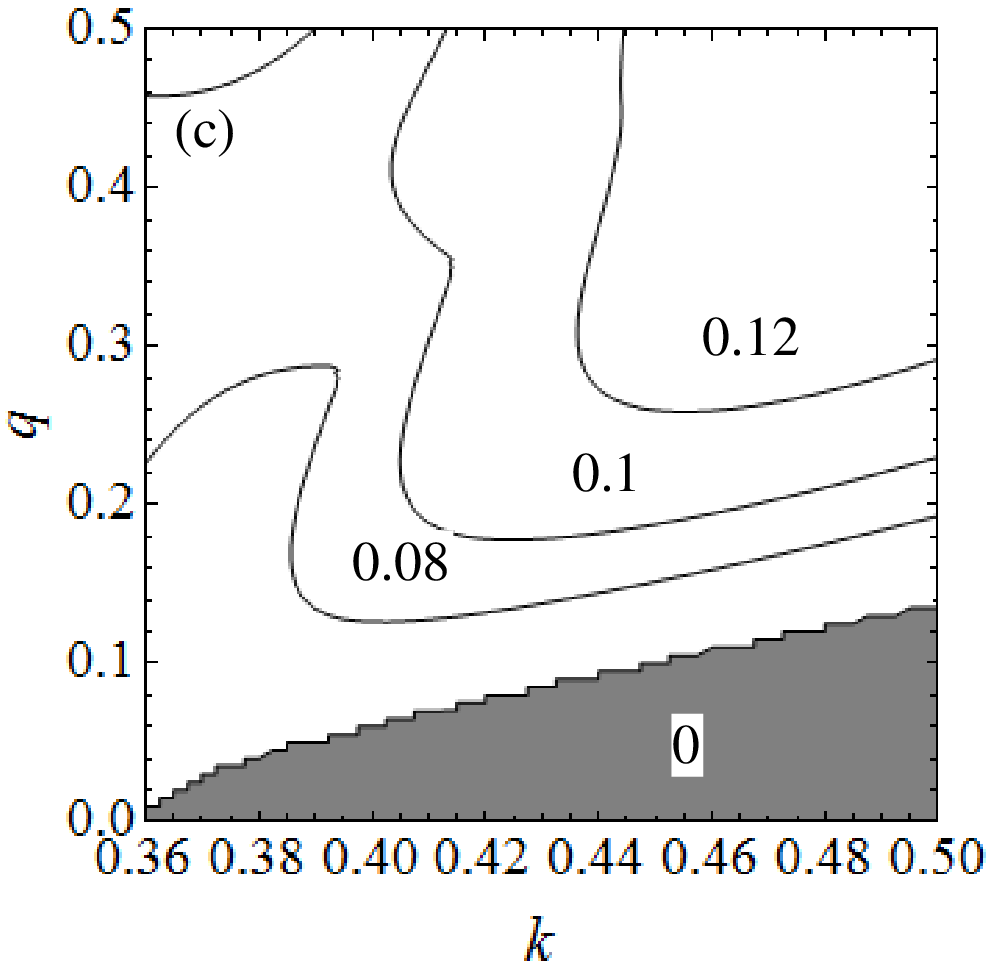}
\includegraphics[scale=.4]{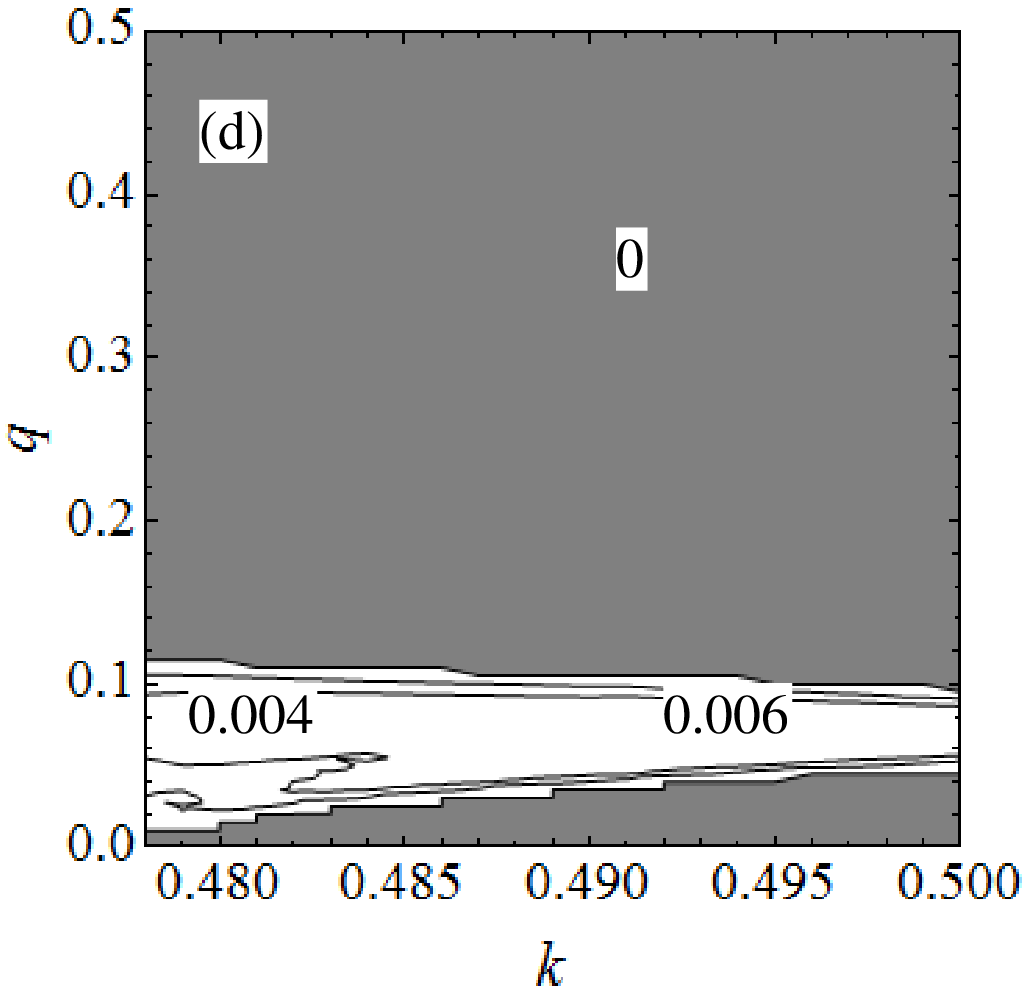}
\caption{The same as Fig.~\ref{c_dynp1} for period-2 solutions for (a) $c_2=0.4$, (b) $c_2=0.1$, (c) $c_2=0.04$, and (d) $c_2=0.01$.
}
\label{c_dynp2}
\end{figure}

For the period-1 case (Fig.~\ref{c_dynp1}), the basic features (dynamically unstable in half the region between the Brillouin-zone center and the zone edge for large $c_2$; the appearance of another instability island near the zone edge and the shrinking of both the unstable domains as the value of $c_2$ is lowered) remain similar to the corresponding situation in the discrete model (Fig.~\ref{d_dynp1}) and also agree with previous results in \cite{wu1}. Similarly, for the period-2 solutions, we find that the plots (Fig.~\ref{c_dynp2}) look quite similar to the corresponding plots from the discrete case (Fig.~\ref{d_dynp2}) up to moderate values of $U$. This again shows that the qualitative features of almost all the properties associated with the continuum model (energy-band structures, stability conditions) can be extracted from the simple discrete model. However, this breaks down when $U$ in the discrete model (or equivalently, $c_2$ in the continuum model) is too large. While in the discrete case, we always find a region of dynamical stability, at large $c_2$ the continuum model has no stable region at all [Fig.~\ref{c_dynp2}(a)]. When we increase $c_2$ gradually from $0.1$ to $0.4$, we notice that the stable region vanishes altogether at $c_2=0.17$, and the instability contours gradually become horizontal. This point will be discussed further in the next section.

\section{The mechanism behind dynamical stability \label{sec:mech}}

We have come across a number of striking features while studying the dynamical stabilities both from the discrete and the continuum models. Here we recall some of them:

1) The period-1 and period-2 states in the lowest energy band are always unstable at $k=0$ for purely sinusoidal modulations with $V_1=0$.
These can, however, be stable for larger $k$ values.

2) In the discrete model with $U\gg K$, the period-2 solutions are more dynamically stable than their period-1 counterparts.

3) In the continuum model the period-2 solutions show greater dynamical stability (compared to the period-1 cases) up to a certain value of $c_2$, but beyond it they become completely unstable. 

In this section we try to explain these features from a physical point of view, and also investigate situations with a non-zero $V_1$ (i.e., a constant component added to the periodic modulation) to obtain a better understanding of the stability mechanism.

The first feature is in complete contrast with BECs in periodic potentials where the $k=0$ state is always dynamically stable. In \cite{wu1}, the dynamical instability of the period-1 Bloch state at $k=0$ for this model with $V_1=0$ was explained in terms of the averaged interaction energy, and it was argued that if the averaged interaction $E_{\rm int} \propto \int_{-p\pi/2}^{p\pi/2}(c_1+c_2 \mbox{cos} 2x) |\psi|^4 d x$ over one period becomes negative, that would make the $k=0$ state unstable. In the case of $V_1=0$ (i.e., $c_1=0$), since the interaction energy (for both period-1 and period-2 solutions) averaged over one supercell is always negative for $k=0$, it resembles a BEC with attractive interparticle interaction, which is unstable dynamically \cite{pethickbook}.

Interestingly, although the lowest Bloch states are dynamically unstable at $k=0$, at larger values of $k$ these can be stable [e.g., the gray-shaded regions in Figs.~\ref{d_dynp1}(a), \ref{d_dynp2}(a), \ref{c_dynp1}(a), and \ref{c_dynp2}(b)]. To explain this seemingly counterintuitive result, we go back to the population density distributions of the discrete model in Figs.~\ref{p1den}, \ref{p2den1}, and \ref{p2den2}. As we have already mentioned, at the zone edge the majority of particles are accumulated in the attractive sites, leaving the repulsive sites nearly empty. Now, for a two-site cell, the transition amplitude between the states with populations $\{|g_1|^2, |g_2|^2\}$ and $\{|g_1|^2\pm 1, |g_2|^2\mp 1\}$ can be estimated as $\sim \sqrt{|g_1| |g_2|} K$. Having alternate empty sites means that the tunneling between neighboring sites is frozen, and the dynamical instability is suppressed. This ``freezing'' takes place for the four-site cell in the case of the period-2 solutions as well. In contrast, at the zone center with $k=0$, the population distribution is more even, and no sites are vacant. The tunneling is non-negligible, and the suppression of dynamical instability does not work around this point. Since the isolation of the higher-density regions, which is responsible for the stability of the superfluid at higher $k$ values, is a result of the attractive interaction in alternate sites, this mechanism can be termed as ``attraction-induced dynamical stability.''

That the period-2 solutions are more stable than the period-1 solutions at higher $U$ values is a direct consequence of the very same mechanism. For period-2 solutions, the higher-density regions are more localized and isolated, i.e., most of the particles are hosted by every fourth site while, for period-1 solutions, it is every second site. In the case of period-1 solutions, this particular stability mechanism is not very prominent near the zone center because the higher-density regions are not separated enough, and a larger $U$ [Fig.~\ref{d_dynp1}(a)] generates more instability than a smaller $U$ [Fig.~\ref{d_dynp1}(b)] for the same value of $k$. On the other hand, for period-2 solutions, a larger $U$ enhances the stability that was already there due to a higher degree of isolation between the higher-density regions. Thus, the superfluid with a higher $U/K$ value [Fig.~\ref{d_dynp2}(a)] is more stable than its lower-$U/K$ counterpart [Fig.~\ref{d_dynp2}(b)] for period-2 solutions.

Of course, there are other factors that determine the dynamical stability, apart from the sign of the net attractive interaction energy, and the suppression of the tunneling due to isolation of higher-density regions. When $U\nu/2K$ is sufficiently small, we observe that a dynamically unstable region appears near the zone edge. This suggests that there are several other factors, too, collectively responsible for the complicated stability diagram like Figs.~\ref{d_dynp1}(b), (c), and (d). It is also worth mentioning here that, similarly, in also BECs in optical lattices with dipole-dipole interactions, it is observed that higher period solutions are more stable \cite{maluckov}.

This attractive-interaction induced dynamical stability is present in the continuum model as well. Only, now the attractive and repulsive ``sites'' are not actual discrete lattice sites any more, but domains. We observe that up to a certain value of $c_2$, increasing the strength of the attractive interaction enhances the stability of the period-2 states around the zone edge by suppressing the inter-site tunneling (Fig.~\ref{c_dynp2}). However, if the nonlinear interaction term is increased even beyond this point ($c_2\simeq 0.17$ here), another mechanism becomes important: the interaction between intra-site particles. Then an increased attractive interaction leads to the collapse of the BEC within a supercell. Since in the discrete model this kind of intra-site degrees of freedom is completely absent, we did not have something equivalent to Fig.~\ref{c_dynp2}(a) there.  

We also note that for higher values of $c_2$ (Fig.~\ref{c_den}) the density distribution has very sharp peaks. As the value of $c_2$ is gradually decreased, those peaks broaden. This is another reason why the discrete model fails to mimic the continuum one for high $c_2$: the expansion of the sharp peaks needs more number of basis functions, and the single-band discrete model is insufficient to capture the actual behavior.

For the excited states, too, there is a departure from the prediction based on the averaged interaction. The period-1 and period-2 states in higher bands usually correspond to an average positive interaction energy, and yet we find the $k=0$ state to be dynamically unstable when $c_1=0$.

\begin{figure}[t!]
\includegraphics[scale=.4]{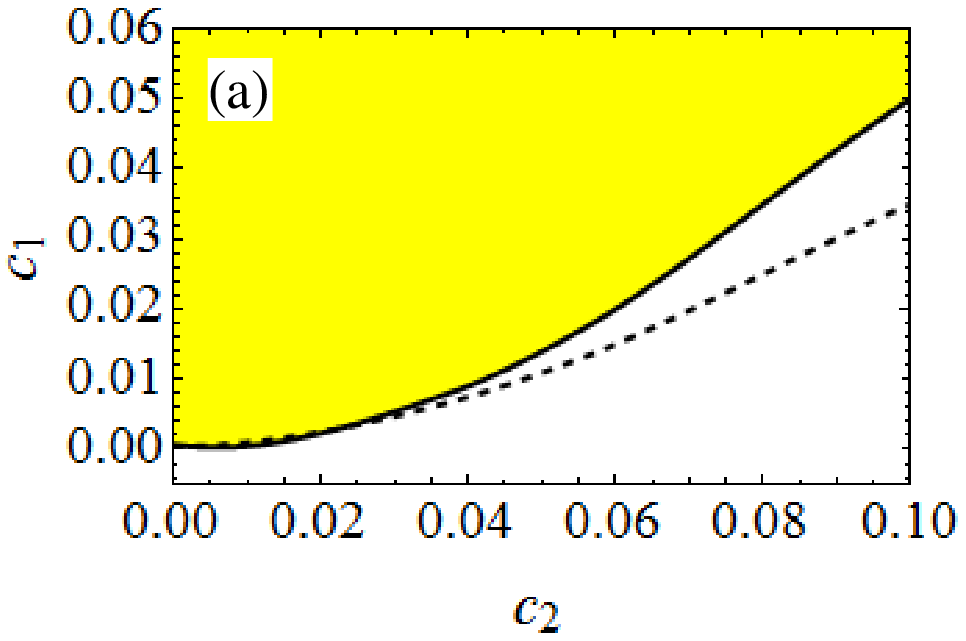}
\includegraphics[scale=.4]{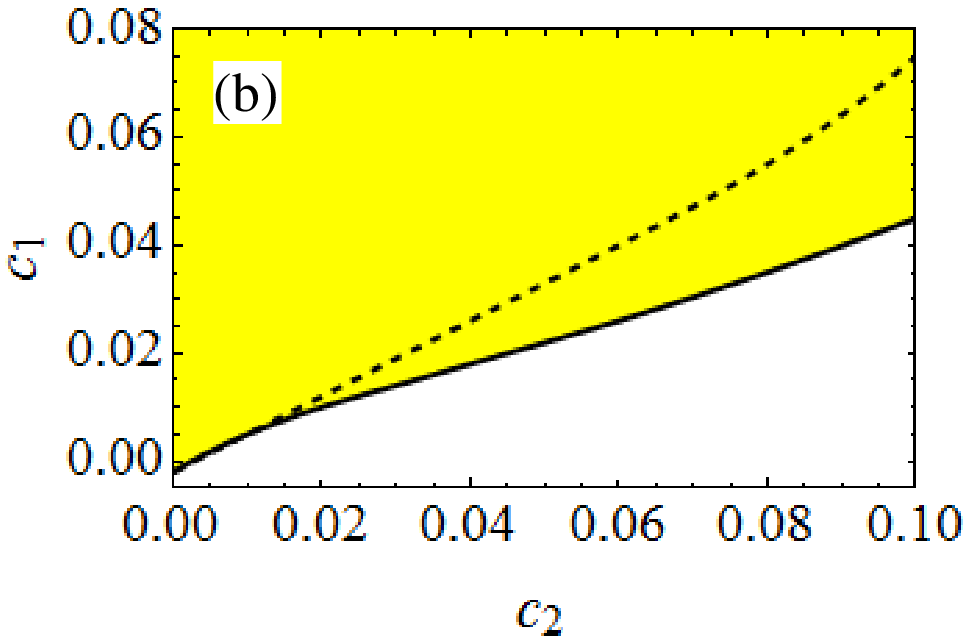}
\includegraphics[scale=.4]{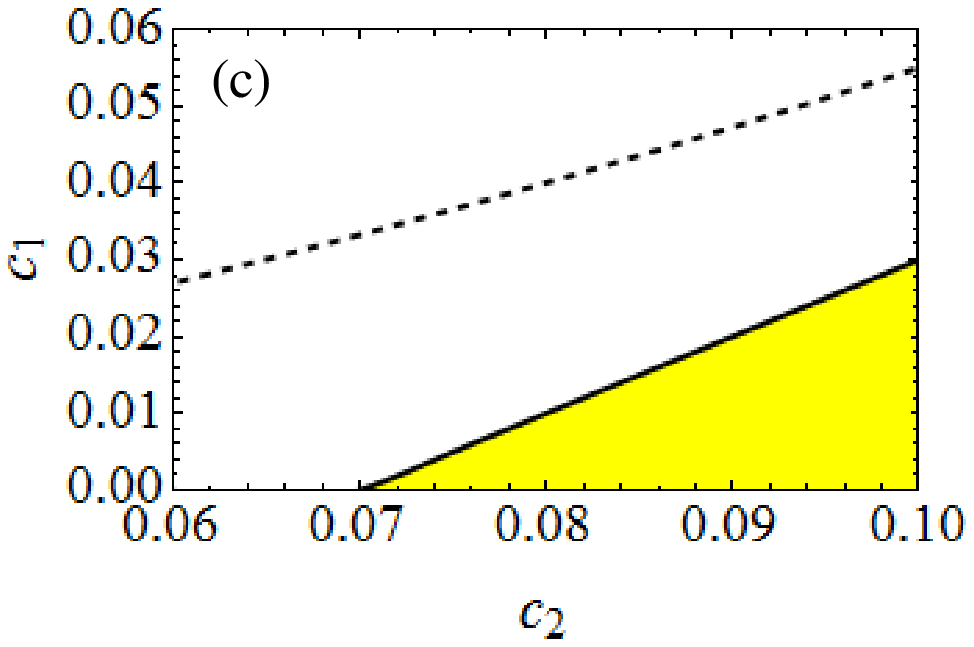}
\caption{(Color online) Dynamical stability and averaged interaction for (a) period-1 and $k=0$, (b) period-1 and $k=1$, and (c) period-2 and $k=0.5$. The dashed lines separate the regions of positive average interaction (above the line) and negative averaged interaction (below the line). The solid line separates the dynamically stable and the unstable regions, and the stable region is shaded in yellow.}
\label{avint}
\end{figure}

Next we consider adding a constant component to the periodic modulation, i.e., taking $c_1\ne 0$ in the continuum model. Although the $k=0$ state in the lowest band is always dynamically unstable for $c_1=0$, by gradually increasing $c_1$ one finally arrives at a critical value that stabilizes the system. In Fig.~\ref{avint}, the solid curve gives the values of the critical $c_1$'s as $c_2$ is increased.
The yellow region bounded by the solid line is dynamically stable, and the white one is dynamically unstable. The dashed line marks the separation between average attractive interaction and average repulsive interaction, i.e., the region below it is attractive and the region above is repulsive. So there is a correspondence between the overall interaction being repulsive, and the system being dynamically stable for period-1 solutions at $k=0$ [Fig. \ref{avint}(a)]. This is in agreement with the results of \cite{wu1}.

In Fig.~\ref{avint}(b), we plot the same for $k=1$ (i.e., the zone boundary for period-1 states). Here, too, there appears to be a relation between the region of dynamical stability and the line where the averaged interaction changes sign. Only, now the solid line lies below the dashed line and the dynamically stable region expands. This can be connected to the ``attractive-interaction induced dynamical stability'' again: near the zone edge there is an additional stability mechanism due to the isolation of the higher-density regions. Thus the system becomes stable even at a $c_1$ value that is slightly lower than the $c_1$ required to make the net interaction repulsive.

The picture, however, changes for period-2 solutions. When $c_2$ is very low, the period-2 branch does not extend up to $k=0$, but rather appears only in a small region around the zone boundary. For a higher value of $c_2$, even though the period-2 branch exists for $k=0$, it is dynamically unstable for $c_1=0$. If we keep on increasing $c_1$, the instability increases. Thus, there is no critical $c_1$ and no stable $k=0$ state for this parameter domain, although the averaged interaction can be both attractive and repulsive, depending on the choices of $c_1$ and $c_2$. 

For period-2 and $k=0.5$ (the zone boundary for period-2 states), the trend is completely opposite to the period-1 results. For $c_2\agt 0.07$, the solutions are dynamically stable even at $c_1=0$, and gradually become dynamically unstable if $c_1$ is increased above a certain value [Fig.~\ref{avint}(c)]. Thus, we have a critical value of $c_1$ that marks the onset of dynamical instability. Below $c_2\simeq 0.07$, the solutions are dynamically unstable at $c_1=0$, and increasing $c_1$ makes it even more unstable. So unlike the period-1 cases, here the dynamically stable region (the region below the solid line, and not above, marked by yellow shading) does not correspond to an overall repulsive interaction [Fig. \ref{avint}(c)].

\begin{figure}[t!]
\includegraphics[scale=.5]{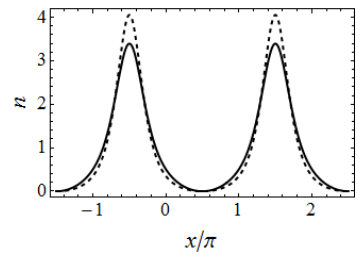}
\caption{Density distributions of period-2 states for $c_2=0.08$ and $k=0.5$ with $c_1=0$ (dashed curve) and $c_1=0.04$ (solid curve). The dashed curve belongs to the stable region and the solid one marks the onset of dynamical instability. Here $x$ is plotted in units of $1/k_0$, $n$ is in units of the average density $n_0$.
}
\label{twopeaks}
\end{figure}

In period-1 situations, the sign of the overall interaction matters in determining the dynamical stability: a repulsive interaction means a dynamically stable BEC. Since the ``attraction-induced dynamical stability'' is not the dominant behavior there (because the higher-density regions are not separated enough), the stability can more or less be accounted for by the sign of the net interaction alone. For period-2 solutions, however, a more complicated factor sets in. Since the period-2 case in general represents a higher degree of isolation between the higher-density regions (Fig.~\ref{c_den}), the tunneling rate here plays a crucial role. For a large $c_2$, the peaks are sharper. The inter-site tunneling is suppressed here and the system is more stable. As $c_2$ is decreased, the peaks spread out to overlap, enabling more tunneling of particles, and that leads to dynamical instability. That is why in Fig.~\ref{avint}(c), the stability region appears in the higher $c_2$ side below the solid line.
That the shape of the peaks and the nature of their separation in the density distribution determines the dynamical stability can be illustrated from Fig.~\ref{twopeaks} as well. The dashed curve of Fig.~\ref{twopeaks} corresponds to the density distribution at $k=0.5$ for $c_2=0.08$ and $c_1=0$, that falls in the stable region of Fig.~\ref{avint}(c). If $c_1$ is increased above $0.04$, although the averaged interaction is now positive (and we could thus expect a stable BEC), we find the region dynamically unstable. Here the density distribution shows wider peaks (the solid curve) and a lesser degree of isolation, and this results in more tunneling of particles, and hence, less stability. So we see that the attractive-interaction induced dynamical stability is the key factor in describing the stability of period-2 states around the zone edge. 

Finally, in a realistic experiment one may anticipate that a harmonic external trapping potential is present in addition to the periodic modulation. In such a trapped case, key modifications would be in the density of states in the low-energy region and the emergence of the quantum pressure due to the inhomogeneity of the system. However, they are relevant only to the long-wavelength perturbations while the fastest growing mode for the dynamical instability in our discussion is the one with a short wavelength of the order of a lattice constant. Therefore, provided the oscillator length of the trap is much larger than the lattice constant, the dynamical stability of the nonlinear lattice in the presence of the harmonic trap could be reliably predicted within the local-density approximation using our results for the untrapped case.

\section{Summary \label{sec:summary}}

We have studied BECs in a nonlinear lattice, i.e., with a spatially periodic scattering length that can be realized via optical Feshbach resonances. Periodic and period-doubled solutions are obtained, both for a reduced discrete model and the full continuum model. The energetic and dynamic stabilities of these stationary states are then examined. It is observed that the periodic nature of the interaction leads to a splitting of the BEC: most of the particles are stored in the attractive sites or domains. If these higher-density regions are not sufficiently isolated and an inter-site tunneling is significant, then the dynamical stability of the superfluid can be qualitatively explained by the sign of the averaged interaction: a net repulsive BEC is stable and a net attractive one is unstable. However, when the higher-density regions are well separated, the inter-site tunneling is suppressed and that enhances the dynamical stability of the system. This ``attraction-induced dynamical stability'' plays the dominant role near the zone edge for periodic solutions. Also, it is this mechanism that renders the higher-periodic solutions more dynamically stable when the nonlinear interaction term is strong enough, unless there is an inter-site dynamics causing a collapse of the BEC.\\

\begin{acknowledgements}
This work was supported by IBS through Project Code (Grant No. IBS-R024-D1); by the Zhejiang University 100 Plan; by the Junior 1000 Talents Plan of China; by the Max Planck Society, the Korea Ministry of Education, Science, and Technology (MEST), Gyeongsangbuk-Do, Pohang City, for the support of JRG at APCTP; and by Basic Science Research Program through National Research Foundation in Korea funded by MEST (Grant No. 2012R1A1A2008028). R.D. would like to acknowledge support from the Department of Science and Technology, Government of India in the form of an Inspire Faculty Award (Grant No. 04/2014/002342). P.V. is supported by the Austrian Federal Ministry of Science, Research, and Economy (BMWFW) and he would also like to thank Prof. Oriol Romero-Isart for support.

\end{acknowledgements}

%\newpage 

\end{document}